\documentclass[runningheads]{llncs}
\usepackage[T1]{fontenc}
\usepackage{pifont}
\usepackage{hyperref}
 \usepackage{tikz}
\usepackage{xcolor}
\usetikzlibrary{positioning,arrows.meta}
\usepackage{tablefootnote}
\usepackage{threeparttable}
\usepackage{makecell}
\usepackage{amsmath}
\usepackage{wasysym}
\usepackage{xcolor}
\usepackage{comment}
\usepackage{graphicx}
\usepackage{amssymb}
\usepackage{booktabs}
\usepackage{pifont}
\usepackage{array}
\usepackage{multirow}
\usepackage{colortbl}
\newcolumntype{C}[1]{>{\centering\arraybackslash}p{#1}}
\usepackage{url}
\usepackage{tabularx}

\usepackage{seqsplit}

\usepackage{xcolor}

\begin{document}
\title{SoK: Privacy-Preserving Order Matching}
\author{Prerna Arote\inst{1} \and
Adrian Saiz\inst{2} \and
Lucianna Kiffer\inst{1}}

\authorrunning{P. Arote et al.}

\institute{IMDEA Networks Institute, Madrid, Spain
\and
École Polytechnique Fédérale de Lausanne (EPFL), Lausanne, Switzerland\\
\email{\{prerna.arote,lucianna.kiffer\}@networks.imdea.org}\\
\email{adrian.saizdepedro@epfl.ch}}

\maketitle              
\begin{abstract}

Privacy-preserving order matching enables traders to discover and execute compatible orders while limiting disclosure of sensitive trading information. Existing systems span financial markets and blockchain-based asset exchanges and employ diverse market structures, privacy mechanisms, and trust assumptions, making their privacy guarantees difficult to compare. We present a systematization of privacy-preserving order matching covering 22 protocol families. We organize the order lifecycle into four phases: order representation and submission, order discovery and collection, order-matching, and settlement, and characterize each phase by protected and exposed information, access roles, privacy mechanisms, and trust assumptions. Our analysis shows that privacy is not a single protocol-level property: protecting selected order attributes or individual phases does not necessarily prevent leakage elsewhere in the lifecycle. We identify recurring trade-offs among privacy, functionality, trust, and verifiability, highlighting the challenge of achieving end-to-end privacy across the complete order lifecycle.

\keywords{Dark pools \and privacy-preserving trading \and secure matching \and order matching}
\end{abstract}

\section{Introduction}
\label{sec:intro}
An executable order reveals a trader's direction, instrument, price, quantity, and timing. On a public (lit) exchange, resting orders are displayed before they trade, exposing trading intent to other market participants. This exposure can be costly for large orders: block trades can move prices, in part before execution, as information about the order leaks~\cite{keim1996upstairs,van2019high}. Dark pools were created to address this problem. A \emph{dark pool} is a trading venue that does not display orders before execution~\cite{petrescu2017dark,sec_regatsn_2018}. Traders submit orders to the operator, which matches them without publicly displaying them, and reports only completed trades. The U.S.\ Securities and Exchange Commission (SEC) describes their purpose as serving investors who want to trade in size while minimizing ``the movement of prices against the trading interest''~\cite{sec2010concept}. Whether dark trading helps or harms market quality remains debated~\cite{zhu2014dark,comerton2015dark,menkveld2017shades,buti2011dark,angel2011equity}; we take no position in that debate.

\paragraph{The operator trust problem.}
A dark pool hides orders from the market, but not from its operator, which must see every 
order in plaintext in order to match it~\cite{ganchev2010censored,eng2013finding}. 
SEC enforcement actions illustrate the consequences of this trust assumption. Operators or their affiliates traded on subscriber orders (at ITG, a secret proprietary desk used live feeds of subscriber orders~\cite{sec_itg_omega_2015}), shared order information with third parties~\cite{sec_ebx_level_2012,sec_creditsuisse_2016}, or failed to restrict internal access~\cite{sec2015ubs,sec_liquidnet_2014,sec_liquidnet_2025}.
The SEC itself acknowledges this risk. In its 2018 rules, it noted ``a risk of information leakage'' of subscribers' confidential trading information to other units of the operator or its affiliates~\cite{sec_regatsn_2018}. Its response was to require operators to disclose their safeguards. Such disclosure increases transparency, but subscribers must still trust the operator to follow its procedures.

\paragraph{From trusting the operator to trusting the protocol.}
Cryptography offers a different answer: reduce the need to trust the operator. The idea is not new. One of the first practical uses of \emph{secure multi-party computation (MPC)} was a sealed-bid auction~\cite{bogetoft2006practical,bogetoft2009secure}, while Thorpe and Parkes proposed securities exchanges in which the public sees only encrypted orders and the operator correct matching~\cite{thorpe2007cryptographic,parkes2015achieving}. Their operator, however, still sees every order. 
Later work removed this point of trust. MPC-based dark pools match \emph{secret-shared} orders without exposing them to any single server~\cite{cartlidge2019mpc,cartlidge2021multi,da2022kicking,da2022all}. Designs based on \emph{homomorphic encryption} let an operator match orders without decrypting them~\cite{asharov2020privacy,balch2020secretmatch,hasselquist2023pet}. \emph{Trusted execution environments (TEEs)} run the matching inside an enclave that the operator cannot inspect~\cite{bentov2019tesseract,tristero}. \emph{Differential privacy} limits what the matching output itself reveals~\cite{chiang2023correlated,polychroniadou2025indifferential}.
These approaches do not remove trust; they move it, to cryptographic assumptions, to a threshold of non-colluding parties, or to a hardware vendor, and some designs still keep a trusted party for part of the process.

Public blockchains change the problem in a different way. They remove the need for a trusted intermediary to hold or settle trades~\cite{notheisen2017trading}, but transactions submitted to public mempools can be visible before execution. This recreates a form of the pre-trade exposure that dark pools were built to avoid, since block producers and other observers can front-run or reorder visible orders~\cite{daian2020flash,eskandari2019sok}.\footnote{A study measured 540.54 million USD of such extractable value, including front-running, over 32 months~\cite{qin2022quantifying}.} Blockchain-native designs therefore combine private computation~\cite{kosba2016hawk} with commitments, zero-knowledge proofs, MPC, or TEEs to build privacy-preserving order-matching systems~\cite{baum2021p2dex,govindarajan2022privacy,garreta2023common,francca2019deep,zhang2019republic,renegade}.

\paragraph{Privacy is not one guarantee.} 
These designs do not provide the same privacy guarantee: each protocol protects specific order information from specific parties, and these choices differ from one design to the next. Some hide orders from the public but not from the operator~\cite{constantinides2021block}; others hide them from each computing party while exposing attributes such as order direction~\cite{cartlidge2021multi}; and some use a delegate that sees orders in plaintext.  So that one party still has direct access to the order~\cite{renegade}.
What is protected also changes over the life of an order. A protocol may hide an order's price and quantity at submission, yet reveal that the order exists or who submitted it. Discovery may reveal which orders are compatible, matching may reveal comparison results, allocation decisions, or partial fills~\cite{da2022all,chiang2023correlated}, and settlement may reveal the resulting trade, for example on a public ledger where deposits and amounts become visible~\cite{baum2021p2dex}. Leakage in any one phase can be actionable even when the order itself stays hidden.
A claim that a protocol is ``privacy-preserving'' is therefore incomplete until it states what is hidden, from whom, and at which stage of the order's life. This motivates examining privacy across the full order lifecycle rather than in the matching computation alone.

\vspace{-2.5pt}
\subsection{This work}
\vspace{-2.5pt}
We systematize privacy-preserving order matching by following the executable order through its life. We model four functional phases: (i)~order representation and submission, (ii)~order discovery and collection, (iii)~order-matching, and (iv)~settlement. 
We include protocols that match two or more orders for financial-market instruments or blockchain-native assets and that use cryptographic, statistical, or trusted-hardware mechanisms to protect information involved in the matching process. 
We exclude generic secure auctions and systems whose privacy mechanisms protect only information outside the matching process, such as transaction transport or post-matching settlement (Section~\ref{sec:corpus}). 
Our corpus comprises 22 protocol families: 13 for financial-market instruments and 9 for blockchain-native assets. We make three contributions.

\medskip
\noindent\textbf{A phase- and role-aware framework.} We characterize privacy by what information is protected or exposed, which roles can access it, the mechanisms used to provide privacy, and the associated trust assumptions.

\medskip
\noindent\textbf{A systematization of 22 protocol families.} We apply this framework to protocols spanning different market structures, privacy mechanisms, and settlement models.

\medskip
\noindent\textbf{Cross-phase findings.} We identify recurring trade-offs and leakage patterns across the order lifecycle, showing that protecting selected order attributes or individual phases does not necessarily prevent information leakage elsewhere in the protocol.

\paragraph{Related surveys.} 
Our work relates to several lines of prior surveys. Surveys of privacy-enhancing technologies cover dark pools only as one application of a broader technique: Baum et al.\ discuss distributed dark pools within finance~\cite{baum2023sok}, Benarroch et al.\ include private limit-order auctions as a use case of programmable privacy~\cite{benarroch2024sok}, and others systematize private computation on blockchains~\cite{almashaqbeh2022sok}, including TEE-assisted~\cite{li2022sok} and FHE-based~\cite{aronoff2025sok} smart contracts. These are organized by technique rather than market function. Surveys of decentralized finance focus on automated market makers and DeFi primitives~\cite{werner2022sok,xu2023sok}. Surveys of front-running and maximal extractable value (MEV) treat hiding an order as a means to prevent reordering~\cite{eskandari2019sok,baum2023mitigation,heimbach2022sok,yang2024sok,passerat2026sok}. In contrast, we treat the order itself as the protected asset, including its exposure to operators and computing parties. Surveys of sealed-bid auctions study one-shot, one-sided mechanisms~\cite{alvarez2020comprehensive}, but order matching is two-sided and repeated. Market microstructure studies analyze dark pools and pre-trade transparency~\cite{angel2011equity,eng2013finding,zhu2014dark}, but take the operator's honesty as given. To our knowledge, no prior SoK is dedicated to privacy-preserving order matching, and none records, for each phase of an order's life, which party can see which order information.

\paragraph{Paper organization.} Section~\ref{sec:background} defines orders, roles, the order lifecycle, and the primitives we consider. Section~\ref{sec:corpus} presents the scope, corpus construction, and classification principles. Sections~\ref{sec:phase1}--\ref{sec:phase4} classify the protocols phase by phase. We conclude with a discussion in Section~\ref{sec:tradeoffs}.

\section{Model and Background}
\label{sec:background}

\subsection{Order Matching}
\label{sec:order-matching}

We use a canonical representation of an order to normalize protocol-specific terminology. Abstractly, an order is
\[
o = (u, a_{\mathrm{in}}, a_{\mathrm{out}}, q, p, \mathit{aux}),
\]
where $u$ denotes the trader, $a_{\mathrm{in}}$ and $a_{\mathrm{out}}$ denote the assets exchanged, $q$ denotes the requested quantity, $p$ denotes the price or price constraint, and $\mathit{aux}$ denotes optional auxiliary information. We refer to everything other than the trader $u$ as the order's \emph{attributes}. This representation is a semantic abstraction rather than a requirement on the concrete format used by a protocol. The direction of the asset pair captures the side of the order: a \emph{sell order} specifies the asset being sold as $a_{\mathrm{in}}$ and the asset being received as $a_{\mathrm{out}}$, while a \emph{buy order} reverses this direction. The price constraint captures the acceptable execution price.

Two orders are compatible when their constraints permit a common execution. For example, a buy order with maximum price $p_b$ and a sell order with minimum price $p_s$ for the same asset are compatible when $p_b \geq p_s$. Matching determines which compatible orders are executed and how the resulting execution is allocated among them. An order may be \emph{fully filled} or \emph{partially filled}, and a single order may be matched against multiple counterparties. Matching may therefore operate on a single pair of orders or jointly consider a larger set, subject to the constraints of the orders and the matching policy. The execution price and allocation rule depend on the market or protocol, and we abstract from these details unless they affect the privacy-preserving matching computation.

\subsection{System Roles}
\label{sec:system-model}

We distinguish five functional roles: 

\medskip
\noindent\textbf{Trader.} A party that creates and submits orders and seeks their execution.

\medskip
\noindent\textbf{Protocol participant.} A party that participates in the protocol or its computation on behalf of itself or the broader protocol, but is not necessarily entrusted with the private information of individual traders. Examples include MPC counterparties or other participants in a distributed matching protocol.

\medskip
\noindent\textbf{Trusted intermediary.} A protocol participant that acts on behalf of a trader and is entrusted by the trader with a function involving its private order information, such as order submission, relay, or matching. Examples include relayers, brokers, and operators. The trust arises from the trader's delegation of this function to the intermediary, and its nature and extent vary across protocols.

\medskip
\noindent\textbf{Settlement mechanism.} The mechanism responsible for executing or recording matched trades, which may be a centralized settlement service, a smart contract, a public blockchain, or another settlement system.

\medskip
\noindent\textbf{Other observer.} A party that can gather information about the protocol or its outcomes without participating in it.

These roles are functional and need not be mutually exclusive: depending on the system architecture, a trader may also participate directly in the protocol, and a party may perform multiple roles. In particular, a protocol participant becomes a trusted intermediary when a trader delegates a function involving its private order information to that party.

\subsection{Order Information and Privacy}
\label{sec:order-information}

Privacy-preserving order matching aims to determine compatible orders and their resulting executions while limiting the disclosure of order information to the parties authorized to learn it. The information that may be protected includes \textbf{order existence}, whether a trader has submitted an order; \textbf{trader--order linkage}, whether an order can be associated with its trader; \textbf{order attributes}; \textbf{matching information}, such as whether two orders are compatible, whether a compatible order exists, or which orders are selected for execution; and \textbf{settlement information}, such as the details of an executed trade and the state that results from it.

Privacy is therefore relative to both the party observing the information and the stage of the order-matching process. A protocol may hide an order attribute from one party while revealing it to another, or protect an order before matching while revealing information about it during matching or settlement. Our taxonomy consequently characterizes privacy separately across the stages of the order-matching lifecycle.

\subsection{Four-Phase Order-Matching Lifecycle}
\label{sec:lifecycle}

\begin{figure}[t!]
\centering
\includegraphics[width=0.85\linewidth]{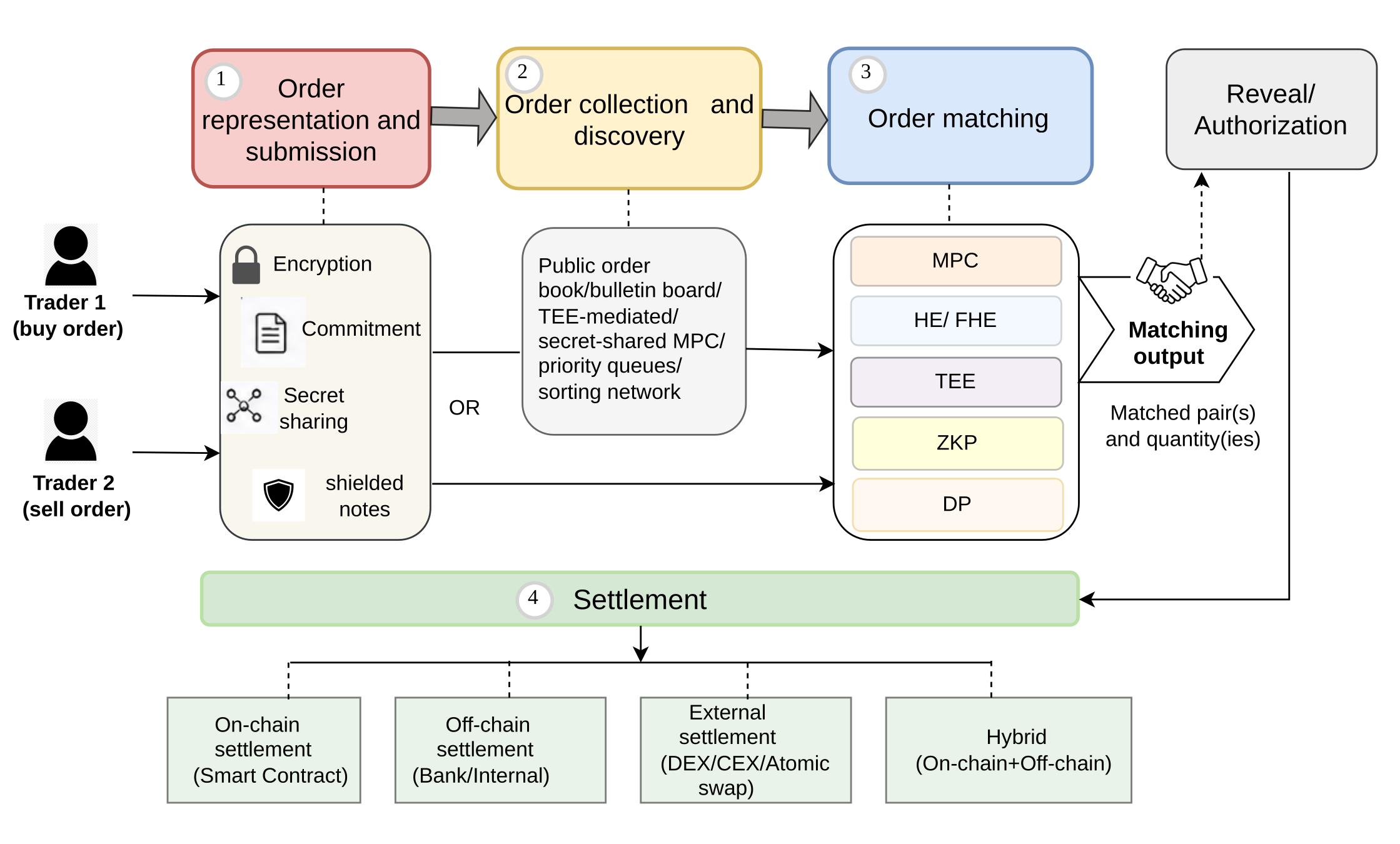}
\caption{Four-phase lifecycle used to analyze privacy-preserving
order matching: order representation and submission, order
discovery/collection, matching, and settlement.} 
\label{fig:ordermatch_lifecycle}
\end{figure}

We analyze each protocol along a common four-phase order-matching lifecycle: \emph{(i) order representation and submission, (ii) order discovery and collection, (iii) order matching, and (iv) settlement}. Figure~\ref{fig:ordermatch_lifecycle} provides a high-level view of this lifecycle. The phases represent functional stages rather than mandatory architectural components; a protocol may combine multiple stages or omit a distinct stage when orders pass directly to the next phase.

\medskip
\noindent\textbf{Phase 1: Order Representation and Submission.} Orders are represented and introduced into the system, determining how they are held and what information is exposed before discovery. We examine order representation, commitment, submission, and pre-discovery privacy.

\medskip
\noindent\textbf{Phase 2: Order Discovery and Collection.} Submitted orders are identified, collected, or organized as inputs to matching. We examine whether discovery identifies counterparties for an individual order or assembles a set of orders, how discovery is organized and timed, and what information it reveals.

\medskip
\noindent\textbf{Phase 3: Order Matching.} The system determines compatible executions and, where necessary, allocates execution across multiple orders. We examine matching scope and policy, the mechanisms and security assumptions used for the computation, and the resulting matching output.

\medskip
\noindent\textbf{Phase 4: Settlement.} Matched orders are converted into an exchange of assets or a corresponding state transition. We examine the settlement mechanism, verification, bindingness, and information exposed by settlement.

Together, these phases provide a common functional decomposition for comparing protocols with substantially different architectures. Section~\ref{sec:corpus} describes how we select and classify the protocols, and Sections~\ref{sec:phase1}--\ref{sec:phase4} examine each phase.

\subsection{Cryptographic Primitives and Security Assumptions}
\label{sec:primitives-security}

Privacy-preserving order-matching protocols rely on cryptographic, statistical, and hardware-based mechanisms to protect order information, perform computations over private data, or verify statements about hidden state. We briefly introduce the mechanisms that recur in our corpus establishing the terminology used in our taxonomy.

\medskip
\noindent\textbf{Secure multi-party computation (MPC)} allows a set of mutually distrusting parties $P_1,P_2,\ldots,P_n$ to jointly evaluate a function $f$ over their private inputs $x_1,\ldots,x_n$, without directly revealing $x_i$ to any other party under the protocol's security assumptions~\cite{cramer2015secure}. In the order-matching protocols we study, however, the MPC output may be revealed to the parties involved, who therefore learn whatever $f(x_1,\ldots,x_n)$ reveals about $x_i$ \cite{renegade}. The practical cost of MPC depends on the complexity of $f$ and the number of participating parties \cite{bogetoft2006practical,bogetoft2009secure}.

\medskip
\noindent\textbf{Homomorphic encryption (HE)} enables computation over ciphertexts without decrypting the underlying plaintexts. Given $c_i=\mathsf{Enc}_{pk}(x_i)$, an evaluator can compute $c_f=\mathsf{Eval}_{pk}(f,c_1,\ldots,c_n)$ such that $\mathsf{Dec}_{sk}(c_f)=f(x_1,\ldots,x_n)$, without learning the individual plaintext inputs during evaluation \cite{yi2014homomorphic}. \emph{Fully homomorphic encryption (FHE)} supports evaluation of arbitrary efficiently computable functions, but typically incurs substantially higher computational and communication costs than plaintext execution \cite{gentry2009fully,balch2020secretmatch}.

\medskip
\noindent\textbf{Cryptographic commitments} bind a party to a value while hiding that value from the verifier \cite{damgaard1998commitment}. A commitment scheme $\mathsf{Com}(x)$ provides \emph{hiding}, which prevents the commitment from revealing $x$, and \emph{binding}, which prevents a commitment from being opened to a different value $x'\neq x$.

\medskip
\noindent\textbf{Zero-knowledge proofs (ZKPs)} enable a prover to convince a verifier that a statement $x$ is valid by demonstrating the existence of a private witness $w$ such that $R(x,w)=1$, without revealing $w$ itself. In order-matching protocols, ZKPs can establish order validity, matching correctness, or the validity of a resulting state transition. They thus provide verifiability and selective disclosure rather than private computation on their own, and are commonly combined with MPC, commitments, or encryption \cite{fiege1987zero,singularity,penumbra}.

\medskip
\noindent\textbf{Differential privacy (DP)} limits what the output of a computation reveals about any individual input by bounding the effect that a small change in the input data can have on the output. Formally, a protocol $\Pi$ is $(\varepsilon,\delta)$-differentially private if, for all input sets $I$ and $I'$ that differ in a single entry and all subsets of outcomes $O$, $\Pr[\Pi(I)\in O] \leq \exp(\varepsilon)\cdot\Pr[\Pi(I')\in O]+\delta$. 

\medskip
\noindent\textbf{Trusted execution environments (TEEs)} provide an isolated execution environment in which code and data are protected from other software running on the host platform \cite{shepherd2024trusted}. In order-matching protocols, the matching procedure can execute within a TEE, keeping order data and intermediate computation state isolated from the surrounding system. TEE-based approaches can run conventional matching algorithms without transforming them into cryptographic protocols, but their security depends on the integrity of trusted hardware, firmware, and attestation mechanisms \cite{khalil2019tex,tristero}.

\paragraph{Security and trust assumptions.}
\label{sec:cryptographic-security}

Security and trust assumptions may apply to the protocol as a whole or to a particular component or phase. For the purposes of this SoK, we abstract away from finer-grained distinctions and classify protocols according to a small set of broad assumption categories that capture the primary sources of their security guarantees. We distinguish \textbf{cryptographic}, when security relies on the assumed hardness of underlying cryptographic problems; \textbf{threshold}, when privacy or security holds provided that no more than a specified number of protocol participants are corrupted or collude; \textbf{hardware}, when security relies on the isolation and integrity of trusted hardware; or \textbf{centralized}, when privacy relies on a designated party, such as a centralized operator or trusted intermediary, not revealing or misusing the information entrusted to it. These categories are not mutually exclusive, and a protocol may rely on multiple types of assumptions.

\section{Systematization Methodology}
\label{sec:corpus}
\subsection{Scope}
\label{sec:scope}

We include protocols that perform matching without specifying or implementing a subsequent settlement mechanism, as well as protocols that integrate matching with internal, custodial, or on-chain settlement.

\medskip
\noindent\textbf{Multi-order matching.} The protocol accepts two or more executable orders, or equivalent order-side inputs, and considers them as potential counterparties or jointly determines trades, allocations, or execution outcomes from the set of orders. Matching may be continuous, periodic, bilateral, or auction-based.

\medskip
\noindent\textbf{Asset-exchange semantics.} The inputs represent the purchase, sale, or exchange of a financial-market instrument or blockchain-native asset.

\medskip
\noindent\textbf{Privacy-preserving matching.} The protocol protects at least one matching-relevant order input, intermediate matching state, or matching output from unauthorized disclosure. Privacy may be provided using cryptographic, statistical, trusted-hardware, or combined mechanisms.

We include protocols that perform matching without specifying or implementing a subsequent settlement mechanism, as well as protocols that integrate matching with internal, custodial, or on-chain settlement.
For blockchain-based protocols, we consider ledgers whose state is publicly verifiable by default, with privacy provided by mechanisms layered on top of the ledger (e.g., commitments, encryption, MPC, or zero-knowledge proofs) rather than by an inherently private ledger.
We exclude generic secure-auction or resource-allocation schemes that do not implement asset-exchange order matching, single-order routing against public liquidity without matching among multiple private orders, and systems whose privacy mechanisms protect only information outside the matching process, such as transaction transport, account state, or post-matching settlement.

\subsection{Corpus Construction}
\label{sec:corpus_construction}

\begin{table}[b]
\centering
\scriptsize
\setlength{\tabcolsep}{4pt}
\renewcommand{\arraystretch}{1.05}
\caption{Corpus construction and screening.}
\label{tab:corpus}
\begin{tabular}{@{}p{2.3cm} p{9.5cm}@{}}
\toprule
\textbf{Dimension} & \textbf{Specification} \\
\midrule

Search sources &
Google Scholar, DBLP, IEEE Xplore, ACM Digital Library, SpringerLink, and publisher/project repositories \\

Search terms &
``privacy-preserving order matching'', ``private order matching'',
``privacy-preserving exchange'', ``private exchange'',
``dark pool'', ``private order book'', ``secure matching'',
``MPC dark pool'', ``private DEX'' \\

Expansion &
Backward and forward citation chaining; project websites and
technical blogs for system identification and implementation details \\

Initial screening &
More than 100 publications and technical sources screened against the
scope and inclusion/exclusion criteria \\

Final corpus &
22 protocol families: 13 financial-instrument and
9 blockchain-based asset. \\
\bottomrule
\end{tabular}
\end{table}

We reviewed more than 100 sources published between 2007 and 2026, including papers from security, privacy, and finance venues (e.g., IEEE S\&P, PETS, NDSS, USENIX Security, and FC), journals, preprints, whitepapers, and technical documentation, identified through keyword searches and citation chaining (Table~\ref{tab:corpus}). Project websites and technical blogs were used primarily for system identification and implementation details.

Applying the criteria in Section~\ref{sec:scope} resulted in a corpus of 22 protocols (Table~\ref{tab:final_corpus} lists the source artifacts identified for each) spanning two broad trading contexts. \emph{Financial-market instrument protocols} (13) primarily process orders for instruments such as equities, bonds, futures, and derivatives. \emph{Blockchain-native asset-exchange protocols} (9) primarily process orders for assets such as cryptocurrencies and tokens whose ownership or state is represented on a blockchain. This distinction describes the trading context of the protocols rather than their privacy architecture or implementation technology.

\subsection{Classification Principles}
\label{sec:classification}

Our classifications record, based on each protocol's sources, which privacy properties it claims to provide and which details it specifies for particular order attributes in 
each phase. We do not independently validate these claims: a claimed guarantee may fail under closer analysis, as shown for Renegade~\cite{arote2026crypto}, which was found not to satisfy some of the privacy guarantees it claims. Such validation or attack is outside the scope of this SoK. We mark a property as ``-'' when the source does not specify it, and do not interpret an unspecified property as a privacy guarantee. We distinguish the mechanisms a protocol uses from the properties they provide, and do not infer a guarantee in one phase from a mechanism used in another. Dimensions such as information exposure and trust assumptions may take multiple values when a protocol combines distinct mechanisms or architectural components; in such cases, we record all explicitly stated classifications. In the following, we classify all 22 protocols along the dimensions of each phase, with a deeper description of each protocol in the Appendix.

\begin{table*}[t]
\centering
\scriptsize
\setlength{\tabcolsep}{2.5pt}
\renewcommand{\arraystretch}{1.05}

\caption{Source artifacts identified (\checkmark); categories are not mutually
exclusive. Protocol type: financial-market (\textcolor{blue!100}{blue}) and blockchain-native (\textcolor{orange!100}{orange}).}

\label{tab:final_corpus}

\resizebox{\textwidth}{!}{%
\begin{tabular}{@{}l*{22}{c}@{}}
\toprule
&
\rotatebox{60}{\textbf{\textcolor{blue!100}{CS}}} &
\rotatebox{60}{\textbf{\textcolor{blue!100}{DPQ}}} &
\rotatebox{60}{\textbf{\textcolor{blue!100}{FM}}} &
\rotatebox{60}{\textbf{\textcolor{blue!100}{IP}}} &
\rotatebox{60}{\textbf{\textcolor{blue!100}{MPC}}} &
\rotatebox{60}{\textbf{\textcolor{blue!100}{PET}}} &
\rotatebox{60}{\textbf{\textcolor{blue!100}{PDP}}} &
\rotatebox{60}{\textbf{\textcolor{blue!100}{AFO}}} &
\rotatebox{60}{\textbf{\textcolor{blue!100}{BA}}} &
\rotatebox{60}{\textbf{\textcolor{blue!100}{CDP}}} &
\rotatebox{60}{\textbf{\textcolor{blue!100}{KB}}} &
\rotatebox{60}{\textbf{\textcolor{blue!100}{TP}}} &
\rotatebox{60}{\textbf{\textcolor{blue!100}{US}}} &
\rotatebox{60}{\textbf{\textcolor{orange!100}{DO}}} &
\rotatebox{60}{\textbf{\textcolor{orange!100}{REP}}} &
\rotatebox{60}{\textbf{\textcolor{orange!100}{SIN}}$^\ast$} &
\rotatebox{60}{\textbf{\textcolor{orange!100}{SPC}}} &
\rotatebox{60}{\textbf{\textcolor{orange!100}{TRI}}} &
\rotatebox{60}{\textbf{\textcolor{orange!100}{COM}}} &
\rotatebox{60}{\textbf{\textcolor{orange!100}{REN}}} &
\rotatebox{60}{\textbf{\textcolor{orange!100}{RIA}}} &
\rotatebox{60}{\textbf{\textcolor{orange!100}{P2D}}}
\\
\midrule

Academic paper
& \checkmark & \checkmark & \checkmark & 
& \checkmark & \checkmark & \checkmark & 
& \checkmark & \checkmark & \checkmark &
\checkmark & \checkmark & & & & \checkmark & & \checkmark & & \checkmark & \checkmark \\

Preprint
& & & & \checkmark & & & & \checkmark & & \checkmark &  & &  &\checkmark & & & & & \checkmark & & & \\

Whitepaper/Technical paper
& & & & & & & & & & & &
& & & \checkmark & \checkmark & & \checkmark & & \checkmark & & \\

Code repository
& & & & & & & & & & & &
& & & \checkmark & \checkmark & & \checkmark  & & \checkmark & & \\

Patent
& & & & & & &  \checkmark& & & & &
& & & & & & & & & & \\

\bottomrule
\end{tabular}%
}

\vspace{2pt}

\parbox{\textwidth}{\scriptsize
\textsuperscript{$\ast$} Portal Gate, the predecessor of Singularity, officially rebranded as Singularity to focus on the institutional crypto market; no formal whitepaper was available during our study~\cite{portalgate,singularity}.\\
\textbf{Protocol key:}
CS = CryptSec~\cite{thorpe2007cryptographic};
DPQ = DI-PQ~\cite{mazloom2023efficient};
FM = FutureMEX~\cite{massacci2018futuresmex};
IP = Indiff-Priv~\cite{polychroniadou2025indifferential};
MPC = MPCJoins~\cite{cartlidge2019mpc};
PET = PETExchange~\cite{hasselquist2023pet};
PDP = PrivacyDP~\cite{asharov2020privacy};
AFO = All-for-One~\cite{da2022all};
BA = BlockAuction~\cite{constantinides2021block};
CDP = CorrDP~\cite{chiang2023correlated};
KB = KickBucket~\cite{da2022kicking};
TP = TurquoisePlato~\cite{cartlidge2021multi};
US = Upending Stock~\cite{jutla2015upending};
DO = DeepOcean~\cite{francca2019deep};
REP = Republic~\cite{zhang2019republic};
SIN = Singularity~\cite{singularity};
SPC = SPCEX~\cite{xie2024spcex};
TRI = Tristero~\cite{tristero,tristero_recent};
COM = COMMON~\cite{garreta2023common};
REN = Renegade~\cite{renegade};
RIA = Rialto~\cite{govindarajan2022privacy};
P2D = P2DEX~\cite{baum2021p2dex}.}

\end{table*}

\section{Phase 1: Order Representation and Submission}
\label{sec:phase1}

Before orders can be discovered and matched, a protocol must represent the trader's intent and make it available to the components that will process it. An order may be represented in plaintext, encrypted, or secret-shared form, and a protocol may additionally commit to the order or to related state so that it can later be verified or used as input to subsequent steps. These choices determine what information is available to the system and what remains hidden before matching. We first characterize how protocols commit to orders and related state, then how orders are submitted and held, and finally what information is exposed before discovery.

\vspace{-6pt}
\subsection{Order Commitment and Submission}
We characterize how orders are committed, submitted, and held before discovery, including the commitment object and mechanism, commitment location, order holder, and submission representation. Protocols may commit to an order itself or to associated state, such as a wallet containing the order. We characterize order commitment along three dimensions:

\begin{table*}[!tb]
\centering
\setlength{\tabcolsep}{3pt}
\caption{Order representation, submission, and pre-discovery exposure in privacy-preserving matching protocols.}  
\vspace{-4pt}
\parbox{\linewidth}{\scriptsize
\textit{Notation:} O = order; S = state; Enc-state = encrypted state; commit = cryptographic commitment; SS-state = secret-shared state;
Both = on-chain + off-chain; T-interm = trusted delegate/intermediary; PP = protocol participants; TEE-infra = TEE infrastructure; PlainTxt = plaintext; SS = secret-shared; -- = not specified;
\textcolor{red}{\ding{51}} = exposed/accessible;
\textcolor{green}{\ding{55}} = not exposed/accessible. \textit{Roles} = parties/observers with access. Order attributes: p = price; q = quantity; pair = asset pair/side; aux = auxiliary attributes.}

\label{tab:phase1-comparison}
\scriptsize
\renewcommand{\arraystretch}{1.08}
\begin{threeparttable}
\resizebox{\textwidth}{!}{%
\begin{tabular}{
p{2.66cm}|
p{1cm}
p{1.55cm}
p{1.15cm}|
p{2.5cm}
p{1.65cm}|
*{2}{>{\centering\arraybackslash}p{1.1cm}}
p{2cm}
p{1.3cm}
}
\toprule
\rowcolor{orange!10}
\textbf{Protocol}
&
\multicolumn{3}{c|}{\textbf{Order Commitment}}
&
\multicolumn{2}{c|}{\textbf{Submission}}
&
\multicolumn{4}{c}{\textbf{Pre-Discovery Privacy}}
\\

\cmidrule(lr){2-4}
\cmidrule(lr){5-6}
\cmidrule(lr){7-10}

&
\textbf{Object}
&
\textbf{Mechanism}
&
\textbf{Location}
&
\textbf{Order Holder}
&
\textbf{Represent.}
&
\textbf{Order--Existence}
&
\textbf{T--O linkage}
&
\textbf{Order Attrib.}
&
\textbf{Role(s)}
\\\hline
\rowcolor{gray!15}
\multicolumn{10}{c}{\textbf{Financial-market instrument protocols}}\\\hline
\textbf{CryptSec} \cite{thorpe2007cryptographic}
& O
& Enc-state 
& Off-chain
& T-interm+ Trader %
& Encrypted
& \textcolor{red}{\ding{51}}  
& \textcolor{red}{\ding{51}} 
& pair 
& T-interm 
\\

\textbf{DI-PQ} \cite{mazloom2023efficient}
& O
& SS-state
& Off-chain
& PP
& SS
& \textcolor{red}{\ding{51}}
& \textcolor{green}{\ding{55}}
& pair
& PP
\\

\textbf{FutureMEX} \cite{massacci2018futuresmex}
& O+S\textsuperscript{$\ast$}
& Commit
& Both 
& Trader+PP
& PlainTxt(O)+ Encrypted(S) 
& \textcolor{red}{\ding{51}}
& \textcolor{green}{\ding{55}}
& p,q aux:time 
& Trader+PP 
\\
\textbf{Indiff-Priv} \cite{polychroniadou2025indifferential}
& O
& Commit 
& Off-chain
& PP 
& Commit 
& \textcolor{red}{\ding{51}} 
& \textcolor{red}{\ding{51}}
& pair, p 
& PP 
\\

\textbf{MPCJoins}\cite{cartlidge2019mpc}
& O+S 
& SS-state
& Off-chain
& PP
& SS
& \textcolor{red}{\ding{51}}
& \textcolor{green}{\ding{55}}
& \textcolor{green}{\ding{55}}
& PP
\\
\textbf{PETExchange} \cite{hasselquist2023pet}
& O
& Enc-state 
& Off-chain
& T-interm 
& Encrypted
&\textcolor{red}{\ding{51}}
& \textcolor{green}{\ding{55}}
& \textcolor{green}{\ding{55}} 
& T-interm 
\\

\textbf{PrivacyDP} \cite{asharov2020privacy,asharov2022methods}
& O
& Enc.state
& Off-chain
& Trader 
& Encrypted
& \textcolor{red}{\ding{51}}
& \textcolor{red}{\ding{51}}
& \textcolor{red}{\ding{51}}
& Trader \\ 


\textbf{All-for-One} \cite{da2022all}
& O
& SS-state 
& Off-chain
& PP\textsuperscript{$\dagger$}
& SS
& \textcolor{red}{\ding{51}} 
& -
& pair, aux:count 
& 
PP
\\

\textbf{BlockAuction} \cite{constantinides2021block}
& O 
& Commit 
& On-chain
& Trader 
& Commit
& \textcolor{red}{\ding{51}} 
& \textcolor{red}{\ding{51}} 
& \textcolor{green}{\ding{55}}
& Trader 
\\

\textbf{Corr-DP} \cite{chiang2023correlated}
& O 
& SS-state
& Off-chain
& PP
& SS
& \textcolor{red}{\ding{51}}
& \textcolor{green}{\ding{55}}
& \textcolor{green}{\ding{55}}
& PP
\\

\textbf{KickBucket} \cite{da2022kicking}
& O
& SS-state
& Off-chain
& PP
& SS
& \textcolor{green}{\ding{55}}
& \textcolor{green}{\ding{55}}
& \textcolor{green}{\ding{55}}
& PP
\\

\textbf{TurquoisePlato} \cite{cartlidge2021multi}
& O
& SS-state
& Off-chain
& PP
& SS
& \textcolor{red}{\ding{51}} 
& \textcolor{green}{\ding{55}}
& \textcolor{green}{\ding{55}}
& T-interm
\\

\textbf{UnpendStock} \cite{jutla2015upending}
& O
& Enc-state 
& Off-chain
& T-interm 
& PlainTxt 
& \textcolor{red}{\ding{51}}
& \textcolor{red}{\ding{51}} 
& p,q,pair
& T-interm 
\\
\hline

\rowcolor{gray!15}
\multicolumn{10}{c}{\textbf{Blockchain-native asset protocols}}\\\hline

\textbf{DeepOcean} \cite{francca2019deep}
& O
& Enc-state 
& Off-chain
& T-interm
& PlainTxt 
& \textcolor{red}{\ding{51}}
& \textcolor{red}{\ding{51}}
& pair, p
& T-interm
\\

\textbf{Republic} \cite{zhang2019republic}
& O
& SS-state
& Off-chain
& PP
& SS
& \textcolor{red}{\ding{51}}
& \textcolor{green}{\ding{55}}
& \textcolor{green}{\ding{55}}
& PP\\

\textbf{Singularity} \cite{singularity,singularity_current}
& O+S
& Enc-state
& Both 
& T-interm
& Commit
& \textcolor{red}{\ding{51}}
& -
& \textcolor{green}{\ding{55}} 
& T-interm
\\
\textbf{SPCEX} \cite{xie2024spcex}
& O
& Enc-state 
& Off-chain 
& T-interm
& Encrypted 
& \textcolor{red}{\ding{51}} 
& \textcolor{green}{\ding{55}}
& pair
& T-interm
\\

\textbf{Tristero} \cite{tristero_recent}
& O
& Enc-state
& Off-chain
& TEE-infra
& Encrypted
& \textcolor{red}{\ding{51}} 
& -
& \textcolor{green}{\ding{55}}
& TEE-infra 
\\


\textbf{Common} \cite{garreta2023common}
& O+S 
& Enc-state 
& Both 
& Trader 
& Encrypted 
& \textcolor{red}{\ding{51}} 
& - 
& pair,aux:max\_p
& Trader+ T-interm 
\\


\textbf{Renegade} \cite{renegade}
& S 
& Commit 
& On-chain  
& T-interm 
& PlainTxt 
& \textcolor{red}{\ding{51}} 
& \textcolor{red}{\ding{51}} 
& O\textsuperscript{$\ddagger$}
& T-interm
\\

\textbf{Rialto} \cite{govindarajan2022privacy}
& S 
& Commit
& Both 
& PP
& SS
& \textcolor{red}{\ding{51}}
& \textcolor{green}{\ding{55}}
& \textcolor{green}{\ding{55}} 
& PP
\\

\textbf{P2DEX} \cite{baum2021p2dex}
& O
& SS-state 
& Off-chain 
& PP 
& SS
& \textcolor{red}{\ding{51}} 
& \textcolor{red}{\ding{51}} 
& pair 
& PP
\\

\bottomrule
\end{tabular}%
}
\begin{minipage}{\textwidth}
\scriptsize
\textsuperscript{$\ast$} Only the trader identifier is committed; round, price index, and volume are broadcast in plaintext. The inventory state is committed in the Merkle tree. \\
\textsuperscript{$\dagger$} No separate operator is assumed; traders jointly perform the MPC computation on their secret-shared orders.\\
\textsuperscript{$\ddagger$} Renegade delegates the order to a relayer (T-interm), which holds it in plaintext; all order attributes are therefore visible to that role.
\end{minipage}
\end{threeparttable}

\end{table*}

\noindent\textbf{Commitment object.} state the protocol commits, with possible options being an \emph{order}, \emph{wallet or related state containing the order}, or \emph{no commitment}.

\noindent\textbf{Committed mechanism.} how an explicitly committed object is represented, with possible options being \emph{cryptographic commitment}, \emph{encrypted state}, or \emph{secret-shared state}.

\noindent\textbf{Commitment location.} where the commitment or committed representation is maintained, with possible options being \emph{on-chain}, \emph{off-chain}, or \emph{both on- and off-chain}.

\noindent\textbf{Order Submission.} We characterize how orders are submitted and held before discovery, including the \textbf{order holder} and the \textbf{submission representation}. The order holder may be the \emph{trader}, a \emph{trusted intermediary}, a \emph{protocol participant}, or \emph{TEE infrastructure}, while the submission may be in \emph{plaintext}, \emph{encrypted}, or \emph{secret-shared} form.

As Table~\ref{tab:phase1-comparison} shows, encrypted state is the most common commitment mechanism (9 protocols), followed by secret-shared state (8), used only with MPC matching~\cite{mazloom2023efficient,cartlidge2019mpc,da2022all,chiang2023correlated,da2022kicking,cartlidge2021multi,zhang2019republic,baum2021p2dex}, and cryptographic commitments (5); six protocols also commit to state beyond the order, such as wallet or balance state. Most (15 of 22) keep commitments off-chain, and on-chain storage does not make orders public, as only commitments, hashes, or ciphertexts are posted~\cite{renegade,garreta2023common,govindarajan2022privacy,tristero}. Commitments bind but do not hide: in CryptSec, BlockAuction, and Renegade, the operator or relayer still sees orders in plaintext~\cite{thorpe2007cryptographic,constantinides2021block,renegade}.

\vspace{-5pt}
\subsection{Pre-Discovery Privacy}

Commitment and submission may protect only selected aspects of an order. We therefore characterize pre-discovery privacy by the order information exposed before discovery or matching:

\textbf{Order existence,} whether a trader has submitted an order; \textbf{trader--order linkage,} whether the contents of an order (e.g., order ID) can be linked to the trader who submitted it; and \textbf{order attributes,} information about the order's parameters, such as asset pair, side, quantity, or price.

For each information type, we record which roles or observers can access it before discovery. Half of the protocols submit orders directly to protocol participants, seven via a trusted intermediary, and only PrivacyDP, BlockAuction, and COMMON leave them with the trader~\cite{asharov2020privacy,constantinides2021block,garreta2023common}. Apart from FutureMEX's public broadcast~\cite{massacci2018futuresmex}, plaintext submission goes only to an intermediary~\cite{jutla2015upending,francca2019deep,renegade}, which sees the full order even when the computing parties do not. Confidential representation does not imply confidential submission: all protocols but KickBucket, which uses dummy orders, expose order existence~\cite{da2022kicking}, and the side (pair) is the most exposed attribute, often while price and quantity stay hidden.

\section{Phase 2: Order Discovery and Collection}
\label{sec:phase2}
After submission, the protocol must determine which orders and, where applicable, which participants form the input to a matching instance, whether by identifying potential counterparties, collecting orders from multiple parties, or organizing already-submitted orders. Protocols differ in who determines these inputs, when, and how the relevant information is exchanged.

\begin{table*}[!tb]
\centering
\setlength{\tabcolsep}{3pt}
\caption{Matching-input discovery before matching, classified by participation,
discovery timing/mechanism, information exposure, and access roles.}
\vspace{-4pt}
\parbox{\linewidth}{\scriptsize
\textit{Notation:}
{\large $\star$}= centralized; \CIRCLE = full participation;
\LEFTcircle = subset participation; Contin. = continuous;
Time-batch = time-based; Event-trigg. = event-triggered;
Central-DA = centralized direct access; Central-PM = centralized
participant-mediated; Distr-broad = distributed broadcast;
Distr-target = distributed targeted query; Distr-other = other;
-- = not specified; (\textcolor{red}{\ding{51}}) = exposed/accessibile;
(\textcolor{green}{\ding{55}}) = not exposed/accessibile.
$p$ = price, $q$ = quantity, pair = asset pair/side,
aux = auxiliary; T-interm = trusted intermediary/delegate,
PP = protocol participants, O-observ = other observers,
CounterP = counterparties.
}
\label{tab:phase2-comparison}
\scriptsize
\renewcommand{\arraystretch}{1.08}

\resizebox{\textwidth}{!}{%
\begin{tabular}{
p{2.66cm}|
p{1.25cm}
p{2.5cm}
p{1.65cm}|
*{4}{>{\centering\arraybackslash}p{1.6cm}}
p{2.5cm}
}
\toprule
\rowcolor{orange!10}
\textbf{Protocol}
&
\multicolumn{3}{c|}{\textbf{Discovery / Collection}}
&
\multicolumn{5}{c}{\textbf{Information Exposure Before Matching}}
\\

\cmidrule(lr){2-4}
\cmidrule(lr){5-9}

&
\textbf{Matching Particip.}
&
\textbf{Discovery Timing}
&
\textbf{Discovery Mechanism}
&
\textbf{Order Existence}
&
\textbf{T--O Linkage}
&
\textbf{Order Attrib.}
&
\textbf{Compatible candidate}
&
\textbf{Role(s)}
\\\hline
\rowcolor{gray!15}
\multicolumn{9}{c}{\textbf{Financial-market instrument protocols}}\\\hline

\textbf{CryptSec} \cite{thorpe2007cryptographic}
& {\Large $\star$} 
& Contin.
& Central\_DA 
& \textcolor{red}{\ding{51}}
& \textcolor{red}{\ding{51}}
& p,q
& \textcolor{green}{\ding{55}}
& PP 
\\

\textbf{DI-PQ} \cite{mazloom2023efficient}
& \CIRCLE
& Contin.
& Central-PM 
& \textcolor{red}{\ding{51}}  
& \textcolor{green}{\ding{55}} 
& \textcolor{green}{\ding{55}} 
& -
& PP
\\

\textbf{FutureMEX} \cite{massacci2018futuresmex}
& \CIRCLE
& Contin. 
& Distr-broad 
& \textcolor{red}{\ding{51}}
& \textcolor{green}{\ding{55}} 
& pair,p,q
& -
& PP
\\

\textbf{Indiff-Priv} \cite{polychroniadou2025indifferential}
& \CIRCLE
& Contin.
& Central-DA
& \textcolor{red}{\ding{51}}
& \textcolor{red}{\ding{51}}
& pair,p,q 
& \textcolor{red}{\ding{51}} 
& PP
\\

\textbf{MPCJoins} \cite{cartlidge2019mpc}
& {\Large $\star$}/ \CIRCLE
& Contin./Time-batch
& Distr-other 
& \textcolor{red}{\ding{51}}
& \textcolor{green}{\ding{55}}
& \textcolor{green}{\ding{55}}
& \textcolor{green}{\ding{55}}
& PP\\


\textbf{PETExchange} \cite{hasselquist2023pet}
& {\Large $\star$} 
& Contin.
& Central\_PM 
& \textcolor{red}{\ding{51}} 
& \textcolor{green}{\ding{55}}
& \textcolor{green}{\ding{55}}
& \textcolor{green}{\ding{55}}

& T-interm 
\\
\textbf{PrivacyDP} \cite{asharov2020privacy,asharov2022methods}
& \CIRCLE
& Contin.
& Central\_DA 
& \textcolor{red}{\ding{51}}
& \textcolor{red}{\ding{51}}
& \textcolor{green}{\ding{55}}
& - 
& T-interm

\\
\textbf{All-for-One} \cite{da2022all}
& \LEFTcircle
& Time-batch
& Distr-broad
& \textcolor{red}{\ding{51}}
& -
& pair
& \textcolor{green}{\ding{55}}
& PP
\\
\textbf{BlockAuction} \cite{constantinides2021block}
& {\Large $\star$}
& Time-batch
& Central-DA
& \textcolor{red}{\ding{51}}
& \textcolor{red}{\ding{51}}
& \textcolor{green}{\ding{55}} 
& \textcolor{green}{\ding{55}}
& PP,O-observ
\\

\textbf{Corr-DP} \cite{chiang2023correlated}
& \LEFTcircle
& Time-batch
& Distr-broad
& \textcolor{red}{\ding{51}}
& \textcolor{green}{\ding{55}}
& \textcolor{green}{\ding{55}}
& -
& PP
\\

\textbf{KickBucket} \cite{da2022kicking}
& \CIRCLE
& Time-batch
& Distr-other 
& \textcolor{green}{\ding{55}}
& \textcolor{green}{\ding{55}}
& \textcolor{green}{\ding{55}}
& \textcolor{green}{\ding{55}}
& PP
\\

\textbf{TurquoisePlato} \cite{cartlidge2021multi}
& \CIRCLE
& Time-batch
& Central-PM 
& \textcolor{red}{\ding{51}}
& \textcolor{green}{\ding{55}}
& \textcolor{green}{\ding{55}}
& -
& T-interm 
\\

\textbf{UnpendStock} \cite{jutla2015upending}
& \LEFTcircle 
& Time-batch
& Distr-broad 
& \textcolor{red}{\ding{51}}
& \textcolor{red}{\ding{51}}
& pair,p,q
& \textcolor{green}{\ding{55}}
& T-interm, PP 
\\\hline

\rowcolor{gray!15}
\multicolumn{9}{c}{\textbf{Blockchain-native asset protocols}}\\\hline


\textbf{DeepOcean} \cite{francca2019deep}
& {\Large $\star$}
& Contin.
& Central-DA
& \textcolor{red}{\ding{51}} 
& \textcolor{red}{\ding{51}} 
& \textcolor{green}{\ding{55}} 
& \textcolor{red}{\ding{51}} 
& CounterP
\\

\textbf{Republic} \cite{zhang2019republic}
& \CIRCLE
& Contin. 
& Distr-broad 
& \textcolor{red}{\ding{51}}
& \textcolor{green}{\ding{55}}
& \textcolor{green}{\ding{55}}
& -
& PP \\

\textbf{Singularity} \cite{singularity,singularity_current}
& {\Large $\star$}
& Contin.
& Central-PM
& \textcolor{red}{\ding{51}} 
& -
& \textcolor{green}{\ding{55}}
& \textcolor{green}{\ding{55}}
& PP\\

\textbf{SPCEX} \cite{xie2024spcex}
& \LEFTcircle
& Contin.
& Central-PM
& \textcolor{red}{\ding{51}} 
& \textcolor{green}{\ding{55}}
& \textcolor{green}{\ding{55}}
& \textcolor{green}{\ding{55}}
& PP
\\

\textbf{Tristero} \cite{tristero_recent}
& \CIRCLE
& Contin.
& Distr-broad
& \textcolor{red}{\ding{51}}
& -
& \textcolor{green}{\ding{55}}
& -
& TEE-infra
\\

\textbf{Common} \cite{garreta2023common}
& {\Large $\star$}
& Time-batch
& Central-DA
& \textcolor{red}{\ding{51}}
& - 
& pair, p
& - 
& T-interm, O-observ 
\\

\textbf{Renegade} \cite{renegade}
& \LEFTcircle  
& Event-trigg.
& Distr-target
& \textcolor{red}{\ding{51}} 
& -
& pair
& \textcolor{red}{\ding{51}} 
& counterP
\\

\textbf{Rialto} \cite{govindarajan2022privacy}
& \CIRCLE
& Event-trigg 
& Central-PM
& - 
& \textcolor{green}{\ding{55}}
& aux:pos 
& \textcolor{green}{\ding{55}}
& PP
\\
\textbf{P2DEX} \cite{baum2021p2dex}
& \CIRCLE
& Event-trigg.
& Distr-other 
& - 
& -
& \textcolor{green}{\ding{55}}
& \textcolor{green}{\ding{55}}
& PP
\\


\bottomrule
\end{tabular}%
}
\end{table*}

Protocols differ in how discovery and the subsequent matching computation are organized across the parties involved. We characterize this organization along two related dimensions: who executes a matching instance and how orders or potential counterparties are collected. These dimensions are independent in general; for example, a protocol may use distributed collection while restricting execution to a subset of protocol participants.

\smallskip
\noindent\textbf{Matching participation.} We distinguish \emph{centralized execution}, in which a designated component executes the matching procedure, from distributed execution. Distributed execution may involve \emph{full participation}, in which all protocol participants execute the matching procedure, or \emph{subset participation}, in which only a subset executes a particular matching instance.

\smallskip
\noindent\textbf{Discovery mechanism.} We distinguish \emph{centralized collection}, in which a designated component collects orders or coordinates their discovery, from \emph{distributed collection}, in which discovery occurs through interactions among protocol participants.
\vspace{-4pt}
\begin{itemize}
    \item \emph{Direct access (Central-DA):} orders are submitted directly to the designated component.
    \item \emph{Participant-mediated (Central-PM):} orders are relayed through trusted intermediaries or other participants.
    \item \emph{Broadcast or dissemination (Distr-broad):} information is distributed to a set of protocol participants.
    \item \emph{Targeted requests or queries (Distr-target):} participants selectively contact potential counterparties or request information from specific parties.
    \item \emph{Other distributed mechanisms (Distr-other):} discovery uses a distributed mechanism that does not fit the preceding categories.
\end{itemize}

As Table~\ref{tab:phase2-comparison} shows, collection is centralized in 12 protocols and distributed in 10. Three protocols do not fit the broadcast or targeted categories and are classified as \emph{Distr-other}. 
In Distr-other protocols, candidate selection instead takes place obliviously inside the MPC computation itself: MPCJoins maintains a secret-shared order book across matching instances~\cite{cartlidge2019mpc}, KickBucket organizes orders into volume buckets within MPC at each clearing point~\cite{da2022kicking}, and P2DEX obliviously sorts orders and discards incompatible ones inside the computation~\cite{baum2021p2dex}. As a result, none of them reveals compatible candidates before matching, whereas candidates are exposed where discovery is an explicit step, driven either by the operator~\cite{polychroniadou2025indifferential,francca2019deep} or by targeted counterparty queries~\cite{renegade}.

\vspace{-5pt}
\subsection{Discovery Timing}

Protocols differ in when orders are collected or potential counterparties are identified relative to the matching process. We distinguish continuous discovery, in which orders may be discovered and incorporated into matching as they arrive, from batched discovery, in which orders are collected  over a defined interval. Some protocols instead trigger discovery in response to a specific event.

\smallskip
\noindent\textbf{Continuous.} Orders are discovered and collected continuously as they are submitted or become available.

\smallskip
\noindent\textbf{Batched.} Orders are collected over a fixed or protocol-defined time interval into discrete matching instances before discovery or matching takes place.

\smallskip
\noindent\textbf{Event-triggered.} Discovery is initiated or advanced in response to a specific protocol event rather than a fixed schedule or continuous arrival of orders.

As Table~\ref{tab:phase2-comparison} shows, continuous discovery is the most common choice (12 protocols) and supports price--time priority, but reveals each order's arrival time. Time-based batching (8 protocols) coarsens this to batch membership and is typical of volume-clearing auctions~\cite{jutla2015upending,chiang2023correlated,da2022kicking,constantinides2021block,garreta2023common}; MPCJoins supports both modes and is counted in each~\cite{cartlidge2019mpc}. Event-triggered discovery appears only in blockchain-native protocols, where matching starts on a protocol event rather than a fixed schedule.

\vspace{-5pt}
\subsection{Discovery Privacy}
Discovery may expose information that was not available to the same parties before discovery began. We therefore characterize discovery privacy using the information categories defined for pre-discovery privacy, augmented with \emph{compatible candidates}, and record which roles or observers can access each category during discovery, including information already exposed at submission.

\noindent\textbf{Order information.} We record whether information about \emph{order existence}, \emph{trader--order linkage}, or \emph{order attributes} is exposed during discovery, and which roles or observers can access it.

\noindent\textbf{Compatible candidates.} We record whether discovery reveals information about potential compatible counterparties or candidate orders, and which roles or observers can access this information.

Table~\ref{tab:phase2-comparison} shows that order existence is exposed during discovery in 19 of 22 protocols; KickBucket is the only protocol that explicitly hides it using dummy orders~\cite{da2022kicking}. Order pair is the most frequently exposed attribute~\cite{massacci2018futuresmex,jutla2015upending,da2022all,polychroniadou2025indifferential,garreta2023common,renegade}, so protocols hiding price and quantity may still reveal direction. Compatible candidates are revealed in only three protocols: the operator selects candidate pairs in Indiff-Priv~\cite{polychroniadou2025indifferential} and DeepOcean~\cite{francca2019deep}, whereas traders discover each other through handshakes in Renegade~\cite{renegade}. In the latter two, a counterparty learns of a compatible order before its value is compared.
\section{Phase 3: Order Matching}
\label{sec:phase3}

Phase 3 describes how submitted orders are evaluated to determine which orders can be matched and, where multiple orders are considered jointly, how the resulting allocation is determined. We characterize the scope and policy of matching, the mechanism and security assumptions used to perform the computation, and the information revealed by the matching output.


\begin{table*}[!tb]
\centering
\setlength{\tabcolsep}{3pt}
\caption{Comparison of matching scope and policy, computation mechanism,
security and trust assumptions, and no-match disclosure.}
\vspace{-4pt}
\parbox{\linewidth}{\scriptsize
\textit{Notation:}
MO-priority = multi-order priority-based;
MO:vol-max = multi-order volume-maximizing;
MO-other = other multi-order policy;
HE = homomorphic encryption;
MPC = secure multi-party computation;
Hybrid = two or more mechanisms;
DP = differential privacy;
TEE = trusted execution environment;
SC = smart contracts;
Crypto. = cryptographic security;
Threshold = threshold security;
central. = centralized security;
H/W-sec = hardware security.
\colorbox{yellow!25}{Yellow} highlights protocols performing pairwise order matching.
}

\label{tab:matching-comparison}
\scriptsize
\renewcommand{\arraystretch}{1.08}

\resizebox{\textwidth}{!}{%
\begin{tabular}{
p{3.5cm}|
p{3.8cm}
p{2cm}
p{2cm}
p{2cm} 
}
\toprule
\rowcolor{orange!10}
\textbf{Protocol}
&
\textbf{Matching Scope and Policy}
&
\textbf{Matching Mechanism}
&
\textbf{Security/ Trust}
&
\textbf{No-match Disclosure}

\\\hline
\rowcolor{gray!15}
\multicolumn{5}{c}{\textbf{Financial-market instrument protocols}}\\\hline

\textbf{CryptSec} \cite{thorpe2007cryptographic}
& MO-priority 
& HE/FHE 
& Hybrid
& Global 
\\

\textbf{DI-PQ} \cite{mazloom2023efficient}
& MO-priority  
& MPC
& Threshold
& None

\\

\textbf{FutureMEX} \cite{massacci2018futuresmex}
& MO-priority 
& MPC
& Crypto 
& Specific 
\\

\textbf{Indiff-Priv} \cite{polychroniadou2025indifferential}
& MO:vol-max
& DP 
& Central.
&
None
\\

\textbf{MPCJoins} \cite{cartlidge2019mpc}
& MO:priority/vol-max/other\textsuperscript{$\dagger$}
& MPC
& Threshold
&
Specific 
\\

\textbf{PETExchange} \cite{hasselquist2023pet}
& MO-priority  
& HE/FHE 
& Hybrid 
& -
\\

\textbf{PrivacyDP} \cite{asharov2020privacy,asharov2022methods}
& - /MO-priority
& FHE/MPC
& Crypto 
& None\\

\textbf{All-for-One} \cite{da2022all}
& MO-priority
& MPC 
& Threshold 
& Specific\\

\textbf{BlockAuction} \cite{constantinides2021block}
& MO:vol-max
& MPC\textsuperscript{4} 
& Hybrid
& Specific
\\

\textbf{Corr-DP} \cite{chiang2023correlated}
& MO: vol-max
& MPC
& Threshold
& None\\

\textbf{KickBucket} \cite{da2022kicking}
& MO:vol-max
& MPC
& Threshold
& None\\

\textbf{TurquoisePlato} \cite{cartlidge2021multi}
& MO-priority 
& MPC
& Threshold  
& None
\\

\textbf{UnpendStock} \cite{jutla2015upending}
& MO:vol-max
& MPC
& Threshold
&
None\\
\hline
\rowcolor{gray!15}
\multicolumn{5}{c}{\textbf{Blockchain-native asset protocols}}\\\hline
\rowcolor{yellow!20}
\textbf{DeepOcean} \cite{francca2019deep}
& Pairwise
& HE\textsuperscript{2} 
& Hybrid 
& Specific
\\

\rowcolor{yellow!20}
\textbf{Republic} \cite{zhang2019republic}
& Pairwise 
& MPC
& Threshold 
& specific 
\\  

\textbf{Singularity} \cite{singularity,singularity_current}
& -
& Hybrid 
& Hybrid 
& -
\\

\textbf{SPCEX} \cite{xie2024spcex}
& MO-Priority
& HE/FHE 
& Hybrid 
& Specific 
\\

\textbf{Tristero} \cite{tristero}
& -
& TEE
& Hardware 
& -

\\
\textbf{Common} \cite{garreta2023common}
& MO:vol-max
& MPC\textsuperscript{1}  
& Hybrid 
& -
\\

\rowcolor{yellow!20}
\textbf{Renegade} \cite{renegade}
& Pairwise
& MPC
& Crypto.
& specific 
\\

\textbf{Rialto} \cite{govindarajan2022privacy}
& MO-priority
& MPC\textsuperscript{3} 
& Central 
& Specific  
\\

\textbf{P2DEX} \cite{baum2021p2dex}
& MO-priority
& MPC
& Threshold 
& None 
\\


\bottomrule
\end{tabular}%
}
\begin{minipage}{\textwidth}
\scriptsize
\textsuperscript{$\dagger$} \emph{MPCjoins}-SVM matches aggregate buy/sell quantities, typically in arrival order, without price comparison \cite{cartlidge2019mpc}.\\
\textsuperscript{1} \emph{COMMON} homomorphically aggregates encrypted order amounts; after decryption, SWAP-ENGINE performs matching and the subsequent Dutch auction in the clear \cite{garreta2023common}.\\
\textsuperscript{2} \emph{Deep Ocean} lets the two traders privately compare order sizes using joint ElGamal encryption, shuffling, blinding, and ZK proofs, revealing only which is larger; the operator then pairs orders using plaintext asset pairs and limit prices \cite{francca2019deep}.\\
\textsuperscript{3} \emph{Rialto} uses MPC to privately sort buy/sell orders by rate, after which the smart contract performs matching in the clear over the ordered list \cite{govindarajan2022privacy}.\\
\textsuperscript{4} \emph{Block Auction} allows MPC or another mechanism for order matching \cite{constantinides2021block}.
\end{minipage}
\end{table*}

\vspace{-5pt}
\subsection{Matching Scope and Policy}

Matching procedures differ in whether they evaluate orders pairwise or consider multiple orders jointly. When multiple orders are considered together, protocols may use different policies to determine which orders are selected and how the resulting allocation is determined. We therefore characterize matching by \emph{scope} and, for multi-order matching, by \emph{allocation policy}.

\smallskip
\noindent\textbf{Matching scope.} We distinguish \emph{pairwise matching}, in which the procedure considers two orders at a time and determines whether they can be matched, whereas \emph{multi-order matching} jointly considers multiple orders to determine an allocation.

\smallskip
\noindent\textbf{Matching policy.} For multi-order matching, we distinguish \emph{priority-based}, in which the allocation follows a priority ordering over orders, such as price or price-time priority; \emph{volume-maximizing}, in which the allocation is chosen to maximize the total quantity traded; and \emph{other}, in which another policy is used to determine the allocation.

Table~\ref{tab:matching-comparison} shows that multi-order matching dominates (16 of 22 protocols), most often with priority-based allocation; five of the six volume-maximizing protocols batch orders by time. MPCJoins is the only \emph{MO-other} protocol: its continuous double auction is priority-based, its periodic auction is volume-maximizing, and its scheduled volume matching fills orders in arrival order until the smaller aggregate volume is exhausted, using an external reference price rather than an internally computed price~\cite{cartlidge2019mpc}. Three pairwise protocols are blockchain-native, simplifying privacy enforcement but providing no global allocation~\cite{francca2019deep,renegade,zhang2019republic}. Matching policy is unspecified for the remaining three protocols~\cite{asharov2020privacy,singularity,tristero}.

\vspace{-5pt}
\subsection{Matching Mechanism and Security}

Protocols employ different mechanisms to perform matching while protecting order information under different trust and security assumptions. We therefore characterize matching by the \emph{mechanism} used for the computation and the \emph{security assumptions} underlying its privacy guarantees.

\noindent\textbf{Matching mechanism.} We distinguish the following mechanisms:
\begin{itemize}
    \item \emph{Secure multi-party computation (MPC):} multiple parties jointly perform the matching computation without revealing their private inputs.
    \item \emph{Homomorphic encryption (HE/FHE):} the matching computation is performed directly on encrypted orders.
    \item \emph{Trusted execution environment (TEE):} the matching computation is performed within a trusted execution environment.
    \item \emph{Differential privacy (DP):} noise or dummy inputs limit what the matching computation reveals about individual orders.
  \end{itemize}

\smallskip
\noindent\textbf{Security assumptions.} We characterize the security assumptions underlying the matching mechanism as \emph{cryptographic (Crypto.)}, \emph{threshold (Threshold)}, \emph{hardware (H/W-sec)}, \emph{centralized (central.)}, or \emph{hybrid}; see Section~\ref{sec:cryptographic-security} for the full definitions.

As Table~\ref{tab:matching-comparison} shows, MPC is the dominant mechanism (14 protocols), and its privacy usually rests on a threshold of non-colluding parties. FutureMEX and Renegade rely on cryptographic assumptions alone, since each party to the computation represents its own order~\cite{massacci2018futuresmex,renegade}. The mechanism alone, however, does not determine what is hidden. Several protocols apply cryptography to only part of the matching and perform the rest in plaintext: Rialto sorts rates in MPC before a smart contract matches the sorted list, COMMON decrypts homomorphically aggregated amounts before matching in the clear, and Deep Ocean protects only the size comparison between two traders~\cite{govindarajan2022privacy,garreta2023common,francca2019deep}. In CryptSec and BlockAuction, the operator decrypts orders and matches them in plaintext, with cryptography serving only verification~\cite{thorpe2007cryptographic,constantinides2021block}. In all these cases, the party executing the plaintext steps sees the values it processes.


\begin{table*}[!tb]
\centering
\setlength{\tabcolsep}{3pt}
\caption{Characterization of matching inputs and outputs and information
exposure across privacy-preserving order-matching protocols.}
\vspace{-4pt}
\parbox{\linewidth}{\scriptsize
\textit{Notation:}
\textcolor{green}{\CIRCLE} = candidate set;
\textcolor{orange}{\CIRCLE} = batch;
\textcolor{blue}{\CIRCLE} = candidate counterparties;
Match-pair = matched pair;
Match-set/alloc = matched set/allocation;
Match-ind = match indicator;
CounterP = counterparties;
PP = protocol participants;
T-interm = trusted intermediary;
Settle-mech = settlement mechanism;
Public = public observer;
Com-proof = commitment/proof;
PlainTxt = plaintext;
SS = secret-shared;
Hybrid = mixed representation;
-- = not specified;
(\textcolor{red}{\ding{51}}) = information exposed/accessibile;
(\textcolor{green}{\ding{55}}) = information not exposed/accessibile.
}
\label{tab:matching-op-comparison}
\scriptsize
\renewcommand{\arraystretch}{1.08}

\resizebox{\textwidth}{!}{%
\begin{tabular}{
p{2.66cm}|
p{1.35cm}
p{3.5cm}
p{1.2cm}
p{1.2cm}
p{1.2cm}
p{2.5cm}
p{3.5cm}
}
\toprule
\rowcolor{orange!10}
\textbf{Protocol}
& \textbf{Matching Inputs}
& \textbf{Matching Output Structure}
& \textbf{Matched quantity}
& \textbf{Clearing price}
& \textbf{CounterP identity}
& \textbf{Matching Output Recipient}
& \textbf{Matching Output Representation}
\\\hline
\rowcolor{gray!15}
\multicolumn{8}{c}{\textbf{Financial-market instrument protocols}}\\\hline

\textbf{CryptSec} \cite{thorpe2007cryptographic}
& \tikz\draw[green,fill=green] (0,0) circle (0.8ex);
& Match-pair
& \textcolor{green}{\ding{55}}
& \textcolor{green}{\ding{55}}
& -
& Public
& Encrypted
\\

\textbf{DI-PQ} \cite{mazloom2023efficient}
& \tikz\draw[orange,fill=orange] (0,0) circle (0.8ex);
& Match-pair
& \textcolor{red}{\ding{51}}
& \textcolor{red}{\ding{51}}
& \textcolor{red}{\ding{51}}
& PP
& PlainTxt
\\

\textbf{FutureMEX} \cite{massacci2018futuresmex}
& \tikz\draw[green,fill=green] (0,0) circle (0.8ex);
& Match-ind
& \textcolor{red}{\ding{51}}
& \textcolor{green}{\ding{55}}
& \textcolor{green}{\ding{55}}
& PP/ Trader
& PlainTxt+Com-proof
\\

\textbf{Indiff-Priv} \cite{polychroniadou2025indifferential}
& \tikz\draw[green,fill=green] (0,0) circle (0.8ex);
& Match-set/alloc
& \textcolor{green}{\ding{55}}
& \textcolor{green}{\ding{55}}
& -
& Trader/PP
& PlainTxt
\\

\textbf{MPCJoins}\textsuperscript{1} \cite{cartlidge2019mpc}
& \tikz\draw[green,fill=green] (0,0) circle (0.8ex); + \tikz\draw[orange,fill=orange] (0,0) circle (0.8ex);
& Match-pair/ Match-set/alloc
& \textcolor{red}{\ding{51}}
& \textcolor{red}{\ding{51}}
& \textcolor{red}{\ding{51}}
& PP
& PlainTxt\textsuperscript{2}
\\

\textbf{PETExchange} \cite{hasselquist2023pet}
& \tikz\draw[green,fill=green] (0,0) circle (0.8ex);
& Match-set/alloc
& \textcolor{red}{\ding{51}}
& \textcolor{green}{\ding{55}}
& -
& PP
& Encrypted
\\

\textbf{PrivacyDP} \cite{asharov2020privacy,asharov2022methods}
& \tikz\draw[green,fill=green] (0,0) circle (0.8ex);
& Match-ind
& \textcolor{green}{\ding{55}}
& \textcolor{green}{\ding{55}}
& -
& PP
& Encrypted
\\

\textbf{All-for-One} \cite{da2022all}
& \tikz\draw[orange,fill=orange] (0,0) circle (0.8ex);
& Match-set/alloc
& \textcolor{red}{\ding{51}}
& \textcolor{green}{\ding{55}}
& -
& PP
& PlainTxt
\\

\textbf{BlockAuction} \cite{constantinides2021block}
& \tikz\draw[orange,fill=orange] (0,0) circle (0.8ex);
& Match-set/alloc
& \textcolor{red}{\ding{51}}
& \textcolor{red}{\ding{51}}
& \textcolor{red}{\ding{51}}
& Public
& PlainTxt+Com-proof
\\

\textbf{Corr-DP} \cite{chiang2023correlated}
& \tikz\draw[orange,fill=orange] (0,0) circle (0.8ex);
& Match-ind
& \textcolor{green}{\ding{55}}
& \textcolor{red}{\ding{51}}
& \textcolor{green}{\ding{55}}
& Trader
& PlainTxt
\\

\textbf{KickBucket} \cite{da2022kicking}
& \tikz\draw[orange,fill=orange] (0,0) circle (0.8ex);
& Match-set/alloc
& \textcolor{red}{\ding{51}}
& \textcolor{green}{\ding{55}}
& -
& PP
& PlainTxt
\\

\textbf{TurquoisePlato} \cite{cartlidge2021multi}
& \tikz\draw[orange,fill=orange] (0,0) circle (0.8ex);
& Match-set/alloc
& \textcolor{red}{\ding{51}}
& \textcolor{red}{\ding{51}}
& \textcolor{red}{\ding{51}}
& PP
& PlainTxt
\\

\textbf{UnpendStock} \cite{jutla2015upending}
& \tikz\draw[orange,fill=orange] (0,0) circle (0.8ex);
& Match-set/alloc
& \textcolor{red}{\ding{51}}
& \textcolor{red}{\ding{51}}
& -
& CounterP
& SS
\\\hline

\rowcolor{gray!15}
\multicolumn{8}{c}{\textbf{Blockchain-native asset protocols}}\\\hline

\textbf{DeepOcean} \cite{francca2019deep}
& \tikz\draw[blue,fill=blue] (0,0) circle (0.8ex);
& Match-pair
& \textcolor{green}{\ding{55}}
& \textcolor{red}{\ding{51}}
& \textcolor{red}{\ding{51}}
& CounterP, T-interm
& Encrypted+PlainTxt(pubkeys)
\\

\textbf{Republic} \cite{zhang2019republic}
& \tikz\draw[blue,fill=blue] (0,0) circle (0.8ex);
& Match-pair
& \textcolor{red}{\ding{51}}
& -
& \textcolor{green}{\ding{55}}
& CounterP, PP
& PlainTxt
\\

\textbf{Singularity} \cite{singularity,singularity_current}
& \tikz\draw[green,fill=green] (0,0) circle (0.8ex);
& Match-ind
& \textcolor{green}{\ding{55}}
& \textcolor{green}{\ding{55}}
& \textcolor{green}{\ding{55}}
& -
& PlainTxt
\\

\textbf{SPCEX} \cite{xie2024spcex}
& \tikz\draw[green,fill=green] (0,0) circle (0.8ex);
& Match-pair
& -
& \textcolor{red}{\ding{51}}
& -
& T-interm, PP
& PlainTxt
\\

\textbf{Tristero} \cite{tristero_recent}
& \tikz\draw[green,fill=green] (0,0) circle (0.8ex);
& -
& \textcolor{red}{\ding{51}}
& -
& \textcolor{red}{\ding{51}}
& Settle-mech
& Encrypted
\\

\textbf{Common} \cite{garreta2023common}
& \tikz\draw[orange,fill=orange] (0,0) circle (0.8ex);
& Match-set/alloc
& \textcolor{red}{\ding{51}}
& \textcolor{red}{\ding{51}}
& \textcolor{green}{\ding{55}}
& Settle-mech
& PlainTxt+Com-proof
\\

\textbf{Renegade} \cite{renegade}
& \tikz\draw[blue,fill=blue] (0,0) circle (0.8ex);
& Match-pair
& \textcolor{green}{\ding{55}}
& \textcolor{red}{\ding{51}}
& \textcolor{red}{\ding{51}}
& CounterP
& SS
\\

\textbf{Rialto} \cite{govindarajan2022privacy}
& \tikz\draw[green,fill=green] (0,0) circle (0.8ex);
& Match-set/alloc
& -
& \textcolor{green}{\ding{55}}
& \textcolor{green}{\ding{55}}
& Settle-mech
& Hybrid\textsuperscript{3}
\\

\textbf{P2DEX} \cite{baum2021p2dex}
& \tikz\draw[green,fill=green] (0,0) circle (0.8ex);
& Match-pair
& \textcolor{green}{\ding{55}}
& \textcolor{red}{\ding{51}}
& \textcolor{red}{\ding{51}}
& CounterP, PP
& PlainTxt
\\

\bottomrule
\end{tabular}%
}
\begin{minipage}{\textwidth}
\scriptsize
\textsuperscript{1} In \emph{MPC Joins}, disclosures depend on the matching mechanism. CDA reveals filled quantities, the resting-order identifier, and execution price; periodic auctions reveal matched-order identities and quantities, followed by the clearing price; scheduled volume matching reveals each order's filled quantity (including zero) and uses an external reference price. Counterparty pairing beyond matched-order identities is unspecified~\cite{cartlidge2019mpc}.\\
\textsuperscript{2} Secret sharing denotes the internal MPC representation, not the final output: computed values are opened to the MPC parties.\\
\textsuperscript{3} Plaintext order indexes and matching decisions are exposed, while rates remain secret-shared and balances are committed and shuffled; the output therefore has mixed protection~\cite{govindarajan2022privacy}.
\end{minipage}
\end{table*}

\vspace{-5pt}
\subsection{Matching Input-Output}

The matching computation operates on different input structures and produces a result that may reveal different information to different parties and in different representations. We therefore characterize matching by its \emph{inputs} and \emph{output}: the matching output by the structure of the result, the information it reveals, the recipient and representation of the result, and the information revealed when no match is found.

\smallskip
\noindent\textbf{Matching inputs.} We distinguish a \emph{candidate set}, where an incoming order is matched against a set of resting orders; a \emph{batch}, where all orders collected in a matching instance are matched jointly; and \emph{candidate counterparties}, where two previously identified counterparties are matched with each other.

\smallskip
\noindent\textbf{Output structure.} We distinguish the following forms of matching output: \emph{Matched pair (Match-pair)---} the output identifies two matched orders; \emph{Matched set/allocation (Match-set/alloc)---} the output identifies multiple matched orders and their allocated quantities;
\emph{Match indicator (Match-ind)---} the output indicates whether a match exists without identifying the matched orders.

\smallskip

\smallskip
\noindent\textbf{Output information.} We record whether the matching output explicitly reveals \emph{matched quantity}, \emph{clearing price}, or \emph{counterparty identity}.

\smallskip
\noindent\textbf{Output recipient.} We record which roles receive the matching result, using the role categories defined in Section~\ref{sec:system-model}.

\smallskip
\noindent\textbf{Output representation.} We characterize the representation in which the matching result is provided as \emph{plaintext (PlainTxt)}, \emph{encrypted}, \emph{secret-shared (SS)}, or \emph{committed/proof-based (Com-proof)}, or \emph{hybrid} when parts of the output use different representations.

\smallskip
\noindent\textbf{No-match disclosure.} We distinguish \emph{specific}, where the protocol reveals that particular orders or counterparties do not match; \emph{global}, where it reveals that no compatible order exists in the relevant set; and \emph{none}, where no explicit no-match result is revealed.

As Table~\ref{tab:matching-op-comparison} shows, even when orders stay hidden during computation, the output is usually opened for execution: 15 protocols deliver it at least partly in plaintext, and in Block Auction it is published to all observers~\cite{constantinides2021block}. Counterparty identity is revealed in eight protocols. A (\textcolor{green}{\ding{55}}) in that column, however, does not mean that matching relationships stay hidden: FutureMEX and Republic reveal matched quantities, COMMON and Corr-DP reveal clearing prices, and in Republic the matched counterparties learn each other's order terms~\cite{massacci2018futuresmex,zhang2019republic,garreta2023common,chiang2023correlated}. Matching can also leak information beyond its intended output, such as the total ordering of rates in Rialto, the relative order of limits in SPCEX, comparison results in PET-Exchange, and which of two orders is larger in Deep Ocean~\cite{govindarajan2022privacy,xie2024spcex,hasselquist2023pet,francca2019deep}. For unmatched orders, nine protocols disclose that specific orders did not match, while eight disclose no explicit no-match result (Table~\ref{tab:matching-comparison}).

\section{Phase 4: Settlement}
\label{sec:phase4}

Phase 4 describes how matched orders are converted into an exchange of assets or a corresponding state transition. We characterize how settlement is performed, the properties that determine how settlement is verified and whether a match commits the parties to settlement, and the information exposed during or after settlement.

\vspace{-5pt}
\subsection{Settlement Mechanism}

Protocols differ in how the exchange resulting from a matched order is executed and recorded. We distinguish settlement performed within the trading system from settlement performed through an external settlement layer, such as a public blockchain or smart contract.

\smallskip
\noindent\textbf{Settlement method.} We distinguish \emph{internal/custodial} settlement, in which the system settles trades by updating balances or positions that it maintains; \emph{on-chain} settlement, in which settlement occurs through a state transition on a public blockchain or smart contract; and \emph{not specified}, when the protocol does not specify how matched trades are settled.

As Table~\ref{tab:settlement} shows, settlement is where the two trading contexts diverge most. Every blockchain-native protocol settles on-chain, whereas 10 of the 13 financial-market protocols leave settlement unspecified, so their privacy guarantees end with the matching output. Only CryptSec and FutureMEX settle internally~\cite{thorpe2007cryptographic,massacci2018futuresmex}, and Block Auction is the only financial-market protocol that settles on-chain, using the blockchain as infrastructure for commitments and result publication rather than as the trading venue~\cite{constantinides2021block}.

\begin{table*}[!tb]
\centering
\setlength{\tabcolsep}{3pt}
\caption{Settlement across selected protocols, characterized by settlement
mechanism, verification method, bindingness, and settlement information
exposure.} 
\vspace{-4pt}
\parbox{\linewidth}{\scriptsize
\textit{Notation:}
-- = not specified;
\textcolor{red}{\ding{51}} = exposed/accessible;
\textcolor{green}{\ding{55}} = not exposed;
Partial = partial information revealed;
Crypto-vrfy = cryptographic verification;
MPC-vrfy = MPC verification;
Settle-val = settlement-layer validation;
TEE-attest = TEE attestation;
None = no explicit verification mechanism specified;
PP = protocol participants;
T-interm = trusted intermediary;
Public = publicly observed/available;
Settle-mech = settlement mechanism;
CounterP = counterparties.
\colorbox{yellow!15}{Yellow} highlights protocols with no settlement mechanism.
}
\label{tab:settlement}
\scriptsize
\renewcommand{\arraystretch}{1.08}

\resizebox{\textwidth}{!}{%
\begin{tabular}{
p{5.05cm}|
p{2.6cm}|
p{2.9cm}|
p{2.3cm}|
p{1cm}
p{1.2cm}
p{1cm}
p{1.1cm}
p{1.5cm}
}
\toprule

\rowcolor{orange!10}
\textbf{Protocol}
&
\textbf{Settle. Method}
&
\textbf{Settle. Verification}
&
\textbf{Settle-Binding}
&
\multicolumn{5}{c}{\textbf{Settlement Privacy}}
\\
\cmidrule{5-9}
&
&
&
&
\textbf{Trade Info}
&
\textbf{T--Trade Linkage}
&
\textbf{Settle State}
&
\textbf{Settle-Activity}
&
\textbf{Role(s)}
\\
\hline

\rowcolor{gray!15}
\multicolumn{9}{c}{\textbf{Financial-market instrument protocols}}
\\
\hline

\textbf{CryptSec} \cite{thorpe2007cryptographic}
& Internal/ custodial 
& Crypto-vrfy
& non-binding 
& \textcolor{green}{\ding{55}}$^\dagger$ 
& - 
& -
& Partial
& Settle-mech \\

\rowcolor{yellow!10}
\textbf{DI-PQ}\cite{mazloom2023efficient}
& -
& -
& -
& - 
& -
& -
& -
& -
\\

\textbf{FutureMEX} \cite{massacci2018futuresmex}
& Internal/ custodial
& Crypto-vrfy+MPC-vrfy
& non-binding
& \textcolor{red}{\ding{51}}
& \textcolor{green}{\ding{55}}
& \textcolor{green}{\ding{55}}
& \textcolor{red}{\ding{51}}
& PP
\\
\rowcolor{yellow!10}
\textbf{Indiff-Priv}\cite{polychroniadou2025indifferential},\textbf{MPCJois}\cite{cartlidge2019mpc}, \textbf{PETExchange}\cite{hasselquist2023pet},  \textbf{PrivacyDP} \cite{asharov2020privacy,asharov2022methods}, \textbf{All-for-One}\cite{da2022all}
& -
& -
& -
& - 
& -
& -
& -
& -
\\

\textbf{BlockAuction}\textsuperscript{$\ast$} \cite{constantinides2021block}
& On-chain & Crypto-vrfy & - & \textcolor{red}{\ding{51}} 
& \textcolor{red}{\ding{51}} & - & - & Public \\
\rowcolor{yellow!10}
\textbf{Corr-DP}\cite{chiang2023correlated},\textbf{KickBucket}\cite{da2022kicking}, \textbf{TurquoisePlato}\cite{cartlidge2021multi}, \textbf{UnpendStock}\cite{jutla2015upending} 
& -
& -
& -
& - 
& -
& -
& -
& -


\\\hline
\rowcolor{gray!15}
\multicolumn{9}{c}{\textbf{Blockchain-native asset protocols}}\\\hline

\textbf{DeepOcean} \cite{francca2019deep}
& On-chain 
& Crypto-vrfy
& -
& \textcolor{red}{\ding{51}} 
& \textcolor{red}{\ding{51}}
& -
& -
& CounterP
\\
\textbf{Republic} \cite{zhang2019republic}
& On-chain
& Crypto-vrfy
& Non-binding
& -
& -
& \textcolor{red}{\ding{51}} 
& \textcolor{red}{\ding{51}} 
& Public 
\\

\textbf{Singularity} \cite{singularity,singularity_current}
& On-chain
& Crypto-vrfy
& Binding
& \textcolor{green}{\ding{55}}
& \textcolor{green}{\ding{55}}
& \textcolor{green}{\ding{55}}
& Partial
& Public
\\
\textbf{SPCEX} \cite{xie2024spcex}
& On-chain
& Crypto-vrfy
& Binding 
& Partial 
& Partial 
& Partial 
& \textcolor{red}{\ding{51}} 
& Public
\\
\textbf{Tristero} \cite{tristero_recent}
& On-chain 
& Crypto-vrfy
& non-binding 
& \textcolor{red}{\ding{51}}
& partial
& \textcolor{red}{\ding{51}}
& \textcolor{red}{\ding{51}}
& public 
\\

\textbf{Common} \cite{garreta2023common}
& On-chain
& Crypto-vrfy
& -
& Partial 
& \textcolor{red}{\ding{51}} 
& partial 
& \textcolor{red}{\ding{51}} 
& Public 
\\

\textbf{Renegade} \cite{renegade}
& On-chain
& Crypto-vrfy
& Non-binding 
& partial 
& \textcolor{red}{\ding{51}}
& \textcolor{green}{\ding{55}}
& \textcolor{red}{\ding{51}}
& CounterP
\\

\textbf{Rialto} \cite{govindarajan2022privacy}
& On-chain
& Crypto-vrfy+MPC-vrfy
& Binding
& Partial 
& Partial 
& Partial 
& \textcolor{red}{\ding{51}} 
& Public
\\


\textbf{P2DEX} \cite{baum2021p2dex}
& On-chain
& MPC-vrfy
& binding 
& \textcolor{red}{\ding{51}} 
& Partial 
& \textcolor{red}{\ding{51}} 
& \textcolor{red}{\ding{51}} 
& Public 
\\

\bottomrule
\end{tabular}}
\begin{minipage}{\textwidth}
\scriptsize
\textsuperscript{$\dagger$} Settlement involves verifying proof corresponding to execution price and numbe of shares; however execution price, traded quantity, and market facts may be selectively disclosed after some delay \cite{thorpe2007cryptographic}.\\
\textsuperscript{$\ast$} BlockAuction is a privacy-preserving double auction for financial dark-pool trading and does not specify the traded assets as blockchain-native assets; the blockchain is used as the protocol infrastructure for commitments, result publication, and verification.
\end{minipage}
\end{table*} 

\vspace{-5pt}
\subsection{Settlement Properties}
\vspace{-1.5pt}
Settlement may provide different guarantees regarding the validity and finality of the resulting trade. We therefore characterize two properties of the settlement process: how the validity of settlement is established and whether a successful match commits the parties to settlement.

\smallskip
\noindent\textbf{Settlement verification.} We characterize how the validity of settlement is established:

\vspace{-10pt}
\begin{itemize}
    \item \emph{Cryptographic verification (Crypto-vrfy):} settlement validity is established using cryptographic commitments or proofs, including zero-knowledge proofs.
    \item \emph{MPC verification (MPC-vrfy):} settlement validity is established through a secure multi-party computation.
    \item \emph{TEE attestation (TEE-attest):} settlement validity is established through attestation of a trusted execution environment.
    \item \emph{None:} the protocol does not use an explicit verification mechanism.
\end{itemize}

\vspace{-5pt}
\noindent\textbf{Settlement bindingness.} We distinguish \emph{binding}, where a successful match commits the parties to settlement, \emph{non-binding}, where matching does not itself guarantee that settlement will occur, and \emph{not specified}, where the protocol does not specify whether matching commits the parties to settlement.

As Table~\ref{tab:settlement} shows, cryptographic verification is used by 11 of the 12 protocols that specify settlement, sometimes combined with MPC as in FutureMEX and Rialto~\cite{massacci2018futuresmex,govindarajan2022privacy}; P2DEX relies on publicly verifiable MPC alone~\cite{baum2021p2dex}. Even Tristero, which matches inside TEEs, binds settlement through a commit--reveal construction rather than enclave attestation~\cite{tristero}. Bindingness varies more: four protocols are binding and five are non-binding. Non-binding designs typically replace enforcement with economic incentives: FutureMEX forfeits the margin deposit of an aborting trader, and Republic relies on bonds because it cannot compel the atomic swap~\cite{massacci2018futuresmex,zhang2019republic}. Binding designs, in turn, still provide refund paths when a party aborts or an order goes unmatched, through escrow timeouts in SPCEX and refunds backed by server collateral in P2DEX~\cite{xie2024spcex,baum2021p2dex}.

\vspace{-5pt}
\subsection{Settlement Privacy}

Settlement may expose information about the matched trade or the resulting state that was not revealed during the preceding phases. We therefore characterize the information exposed during or after settlement and the roles or observers that can access it.

\smallskip

\noindent\textbf{Information exposure.} For each information type, we record whether it is \emph{exposed}, \emph{not exposed}, \emph{partially exposed}, or \emph{not specified}: \emph{trade information (Trade Info)}, the settled asset/instrument, quantity, or price; \emph{trader--trade linkage (T--Trade Linkage)}, whether a settled trade can be linked to its traders; \emph{settlement state (Settle State)}, resulting balances, positions, or other state; and \emph{settlement activity (Settle-Activity)}, settlement execution, timing, or related activity.

For each information type, we also record which roles or observers can access it during or after settlement; see Section~\ref{sec:system-model} for the role definitions.\\
As Table~\ref{tab:settlement} shows, settlement shifts the privacy boundary toward the public: in eight of the 12 protocols that specify settlement, settlement information is accessible to all observers, and settlement activity is exposed in eight. Singularity is the only protocol that hides trade information, trader--trade linkage, and settlement state on chain, publishing only nullifiers and new note commitments~\cite{singularity_current}. Several protocols expose only partial information by design: Rialto opens only the aggregate fee across matched pairs, COMMON reveals the pair and round but not individual amounts, and SPCEX exposes escrow transfers while mixing unlinks them from traders~\cite{govindarajan2022privacy,garreta2023common,xie2024spcex}. Such mitigations target linkage rather than activity, so even protocols that hide orders throughout matching reveal that a settlement took place.

\section{Trade-offs and Discussion}
\label{sec:tradeoffs}

Our taxonomy shows that privacy is not a single protocol-level property: it depends on the information, observer, and lifecycle phase.

\subsection{Privacy Across the Order Lifecycle}
\label{sec:lifecycle-privacy}
\vspace{-4pt}
\noindent\textbf{Phase- and information-dependent privacy.}
No protocol uniformly protects an order from submission through settlement. Information protected at one phase is often exposed later: 15 protocols expose matching outputs at least partly in plaintext, while 8 of 12 protocols specifying settlement expose settlement information publicly. Protection varies by information type: price and quantity are often hidden, but order existence is exposed in all but KickBucket~\cite{da2022kicking}, and order side is the most frequently exposed attribute. Settlement may further expose activity or enable linkage even when order contents remain hidden.

\smallskip
\noindent\textbf{Discovery and matching leakage.}
Explicit candidate selection by an operator~\cite{polychroniadou2025indifferential,francca2019deep} or through trader handshakes~\cite{renegade} reveals compatibility before order values are compared, whereas oblivious MPC-based selection avoids this~\cite{cartlidge2019mpc,da2022kicking,baum2021p2dex}. Matching may also leak beyond its intended output, e.g., rate ordering in Rialto~\cite{govindarajan2022privacy} and relative limit ordering in SPCEX~\cite{xie2024spcex}. Thus, ``private matching'' alone does not characterize end-to-end privacy.

\subsection{Privacy, Functionality, and Trust}
\label{sec:privacy-functionality-trust}
\vspace{-4pt}
\noindent\textbf{Privacy--functionality.}
Pairwise matching, used by three blockchain-native protocols, simplifies private computation but provides no global allocation; multi-order matching, used by 16 protocols, requires more complex joint computation. Richer functionalities often reintroduces leakage: continuous discovery enables price--time priority but reveals arrival times, while time-based batching coarsens this to batch membership at the cost of latency.

\smallskip
\noindent\textbf{Privacy--trust.}
Privacy mechanisms shift trust. MPC, used by 14 protocols, replaces a trusted operator with a threshold assumption on non-colluding parties; TEE-based designs rely on trusted hardware~\cite{tristero_recent}. Seven protocols use a trusted intermediary, three of which receive orders in plaintext~\cite{jutla2015upending,francca2019deep,renegade}. Where each party contributes only its own input, as in FutureMEX and Renegade, privacy can instead rest on cryptographic assumptions alone~\cite{massacci2018futuresmex,renegade}.

\smallskip
\noindent\textbf{Privacy--verifiability.}
Verification typically requires disclosure. Protocols that settle on-chain make execution publicly checkable precisely by exposing it, and those that limit exposure do so by deciding what to reveal: Rialto opens only the aggregate fee across matched pairs~\cite{govindarajan2022privacy}, and COMMON reveals the pair and round but not individual amounts~\cite{garreta2023common}. Only Singularity hides trade information, linkage, and settlement state on chain~\cite{singularity}.

\subsection{Toward End-to-End Privacy Guarantees}
\vspace{-4pt}
Phase-level guarantees do not automatically compose: information hidden during matching may become linkable through discovery, matching output, or settlement. End-to-end privacy therefore requires tracking cumulative leakage across all phases and specifying \emph{(i)} the protected information, \emph{(ii)} the observer, and \emph{(iii)} the phase, while distinguishing intended from incidental disclosures. Our corpus highlights a reporting gap: settlement information is unspecified for 10 of 13 financial-market protocols. A standardized framework covering these dimensions, representation, and trust assumptions would enable precise and comparable privacy claims.

\bibliographystyle{splncs04}
\bibliography{references}
\appendix
\section{Protocol Profiles}
\label{sec:protocol-profiles}

The appendix provides detailed profiles of the protocols included in our
corpus. Within each primary trading-object category, we assign protocols to
broad design families based on their dominant architectural or technical
characteristics. These families are used to organize the presentation and
do not constitute an additional classification dimension. A protocol may
use techniques associated with multiple families; the four-phase
classification remains the basis of the main comparison.

We use the following notation for the roles defined in Section~\ref{sec:system-model}: \textit{Trader}, protocol participant (\textit{PP}), settlement mechanism (\textit{Settle-mechanism}), other observers (\textit{O-observ}), and trusted delegate or intermediary (\textit{T-interm}).

\subsection{Financial-Market Protocols}
\label{app:financial-protocols}

The financial-market protocols are grouped into \emph{Secure Financial-Market and Auction Designs}, \emph{MPC-Based Dark-Pool and Matching Designs}, and \emph{Encrypted Matching and Controlled-Disclosure Designs}.

\subsubsection{\S \; Secure Financial-Market and Auction Designs}\mbox{}\par

\medskip
\noindent\textbf{CryptSec \cite{thorpe2007cryptographic}}\\
\noindent\emph{Design goal.} A continuous double-auction exchange in which the operator hides order values from the market while publicly proving that every match follows the published rules.

\noindent\emph{Entities.} Traders, market operator (PP), third-party bulletin board (T-interm, Settle-mechanism), clearing agent (PP).

\noindent\emph{Workflow.} A trader encrypts price and quantity under the operator's Paillier key and posts them, with the order side, to an independent bulletin board that timestamps, publishes, and signs a receipt. The trader privately sends the operator the encryption randomness, which proves knowledge of the plaintext and prevents orders derived from another trader's ciphertext. The operator decrypts, matches in plaintext under price--time priority, logs each trade, and republishes the encrypted order book with range proofs for every match and order rank. The clearing agent receives the randomness needed to verify execution price and quantity, and trade is finally settled by bulletin board.

\noindent\emph{Privacy profile.} Price and quantity enter encrypted, while side, timestamp, and order existence are public; trader anonymity is out of scope. The operator decrypts every order, so confidentiality holds only against traders and the public. They see ciphertexts, matched order identifiers, and proofs that leak the book's relative price ranking. Execution price and quantity reach the clearing agent in plaintext and are published only after a designer-chosen delay.\\

\noindent\textbf{FuturesMEX \cite{massacci2018futuresmex}\\}
\noindent\emph{Design goal.} A fully distributed futures exchange with limit and market orders, margin accounting, and mark-to-market, run without a trusted operator. Traders stay anonymous and their inventories confidential, to prevent price discrimination.

\noindent\emph{Entities.} Traders (PP), BFT distributed ledger for broadcast, anonymous network, private payment system (Settle-mechanism).

\noindent\emph{Workflow.} Traders deposit cash from a private payment system and commit to an inventory token (cash, contracts, pending-order estimates, solvency flags) stored as a Merkle-tree leaf. To post an order, a trader broadcasts its price level, signed volume, and a commitment to its identity. It spends its old token and proves in zero knowledge that its updated inventory stays solvent. Matching runs in the clear on the public book, against the earliest opposite order at the same price; each matched trader proves order ownership and its inventory update in zero knowledge. Since any order can change others' solvency, all traders re-prove their net position after every update, and an MPC check (SPDZ) detects newly insolvent traders, who must cancel and liquidate. At the final round, positions are marked to market at the mid-price.

\smallskip
\noindent\colorbox{gray!12}
{\parbox{\dimexpr\linewidth-2\fboxsep\relax}{%
\textbf{Penalty mechanism.}
Each trader's initial margin is frozen as a deposit. A trader that aborts a protocol step is identified, because honest traders must prove participation in that round, and it cannot reclaim its deposit at mark-to-market. Its funds can then be redistributed among the honest traders.
}}\\

\noindent\emph{Privacy profile.} Orders enter the public book in plaintext by design: price, volume, side, and time. They are unlinkable to their trader through committed identities, anonymous networking, and one-time inventory tokens. Cash, positions, and net positions remain committed throughout. Matching outputs are public plaintext notifications of the matched order, price, and volume, and counterparties never learn each other's identity. Privacy therefore protects who trades and what they hold, not what is traded.\\

\noindent\textbf{Upending Stock \cite{jutla2015upending}\\}
\noindent\emph{Design goal.} Replace the continuous double auction with a periodic sealed-bid clearing-price auction whose auctioneer is emulated by MPC, so that no party or small coalition learns the order book.

\noindent\emph{Entities.} Clients (Trader), brokers (T-interm), five MPC parties such as four brokers and a regulator (PP), public bulletin board, settlement authority (Settle-mechanism).

\noindent\emph{Workflow.} Clients send market or limit orders (a side and a list of quantity--price pairs) to their broker. The broker tags each order with its broker ID and a random order ID, then encrypts or secret-shares it among the MPC parties. At the close of each 15--30-minute session, the MPC builds cumulative demand and supply curves and sets a single clearing price where they cross. Market orders and limit orders beyond that price fill fully, and the residual at the marginal price goes largest-first, leaving at most one partial fill. Results are posted on the bulletin board and an external authority transfers the shares. Transcripts are released for audit about a month later, which is the paper's substitute for malicious security.

\noindent\emph{Privacy profile.} Brokers receive their clients' orders in plaintext; the MPC parties see only encrypted or secret-shared inputs, under semi-honest non-collusion. The public plaintext output is the order count per side, the clearing price, and the broker and order IDs of filled orders with their allocations. A variant publishes only price and volume and sends allocations encrypted to each broker. Unfilled orders are never revealed, although the delayed audit transcripts may eventually expose inputs.\\

\noindent\textbf{BlockAuction \cite{constantinides2021block}\\} 
\noindent\emph{Design goal.} A periodic double auction for dark-pool trading in which orders stay hidden until the auction and a smart-contract audit trail lets anyone verify the operator's results.

\noindent\emph{Entities.} Clients (Trader), auction operator (PP), smart contract used as a public bulletin board.

\noindent\emph{Workflow.} The operator registers each client address with a client-specific public key. A client encrypts its order (side, asset, limit price, volume, minimum execution size) with a random nonce and posts a hash of the ciphertext, plus a deposit, to the contract. In the reveal phase it posts the ciphertext, which anyone can check against the commitment. The operator decrypts all orders and matches them off-chain under a publicly announced clearing algorithm, then publishes each match with the decryption keys of the matched orders. Clients verify the results collectively and can raise fraud claims on-chain, where the record serves as legal evidence. Settlement is outside the protocol.

\medskip
\noindent\colorbox{gray!12}
{\parbox{\dimexpr\linewidth-2\fboxsep\relax}{%
\textbf{Penalty mechanism.}
The deposit posted with each commitment is returned only when the client reveals its ciphertext. This also makes it costly for the operator to inject fake clients and reveal only the orders that benefit it.
}}\\

\noindent\emph{Privacy profile.} Orders enter as hash commitments, so before the reveal only the client's address and participation are public, not even the asset. From the reveal onward the operator sees every order in plaintext, so confidentiality rests on trusting it. The output is public plaintext: matched orders, with their counterparties and terms, are visible to everyone. Unmatched orders remain known only to their owner and the operator.

\subsubsection{\S \; MPC-Based Dark-Pool and Matching Designs} \mbox{}\par

\medskip
\noindent\textbf{MPCJoins \cite{cartlidge2019mpc} \\} 
\noindent\emph{Design goal.} Replace the trusted dark-pool operator with MPC servers, and measure which standard matching mechanisms are fast enough to run under MPC.

\noindent\emph{Entities.} Traders; two (full-threshold SPDZ) or three (honest-majority Shamir) MPC servers, such as the venue operator and a regulator (PP).

\noindent\emph{Workflow.} A trader masks each order value with a random value that the servers open to it, so the servers end up holding the order's owner, volume, and price secret-shared in separate buy and sell lists. Three mechanisms are implemented:
\begin{itemize}
  \item \emph{Continuous double auction.} An incoming order is matched against the private sorted book under price--time priority, at the resting order's price; any remainder is inserted obliviously.
  \item \emph{Periodic auction.} Orders are inserted into sorted lists during the period. At close, the servers match them and compute the volume-maximizing clearing price.
  \item \emph{Scheduled volume matching.} Only volumes are submitted. Orders fill in arrival order up to the smaller aggregate side, at an external reference price.
\end{itemize}
The resulting matching outputs are returned to the relevant traders or protocol participants. The paper does not define a complete cryptographic asset-transfer or securities-settlement procedure. 

\noindent\emph{Privacy profile.} Price, volume, and owner remain secret-shared, secure with abort against active adversaries unless both of two, or two of three, servers collude. The order side leaks through the separate buy and sell lists. Outputs are opened in plaintext to the servers, and differ by mechanism:
\begin{itemize}
  \item \emph{Continuous double auction:} each execution's resting-order owner, quantity, and price.
  \item \emph{Periodic auction:} the matched owners and quantities, then the clearing price.
  \item \emph{Volume matching:} every order's filled quantity, zero for unfilled orders.
\end{itemize}
Unmatched orders stay hidden in the first two mechanisms.\\

\noindent\textbf{KickBucket \cite{da2022kicking}}\par
\noindent\emph{Design goal.} Fast periodic dark-pool trading of one instrument at an external reference price. MPC servers replace the operator, and orders stay hidden from them, including their direction and whether an investor trades at all.

\noindent\emph{Entities.} Investors (Traders), three MPC servers emulating the operator (PP).

\noindent\emph{Workflow.} Investors secret-share each order among the servers: an identity, two direction bits (neither set marks a dummy order), and, in volume match, a bit-decomposed volume. A randomized check rejects malformed inputs. In bucket match every order has a fixed volume, so large orders are split; bucket-2 adds a cross-list clearing between two bucket sizes. At each scheduled cross, the servers open which side has more total volume, open the smaller side's direction bits, and fill those orders completely. The larger side is then opened in arrival order until the smaller side is covered. In volume match, the last order's remainder is deducted without opening its volume.

\noindent\emph{Privacy profile.} Orders enter secret-shared, with direction and volume hidden, under an honest-majority assumption with security-with-abort. Dummy orders hide whether an investor trades at all. The servers learn in plaintext which side is larger and the identities and volumes of filled orders. Opening direction bits also narrows which remaining orders are real, which dummies dilute. Bucket-2 may leak a partially filled order's remainder; volume match only leaks that its owner may still hold volume.

\newpage
\noindent\textbf{Turquoise Plato \cite{cartlidge2021multi}}\par
\noindent\emph{Design goal.} An MPC emulation of LSEG's Turquoise Plato Uncross dark pool across thousands of instruments, so that no hosting organisation can see or misuse orders, including which instrument they trade.

\noindent\emph{Entities.} Traders; two (full-threshold SPDZ) or three (honest-majority Shamir) organisations jointly running a gateway engine and several auction engines (PP).

\noindent\emph{Workflow.} Instruments are obliviously assigned to auction engines through a Paillier-encrypted lookup vector. A trader picks its instrument's entry, re-randomises it, and sends its secret-shared order (name, instrument, side, volume, minimum execution size) to the gateway. The gateway decrypts only the destination engine. Orders arriving in each five-second interval are inserted obliviously into buy and sell lists sorted by size, then time. They are uncrossed in the next interval at the primary exchange's midpoint, subject to minimum execution sizes, and the book is flushed so that unfilled orders must be re-entered. No settlement is specified.

\noindent\emph{Privacy profile.} Name, volume, minimum execution size, and instrument enter secret-shared, while the side is known to the servers through the buy and sell lists. Routing reveals which subset of instruments an order belongs to. At uncrossing, the servers learn in plaintext each trade's counterparty names, instrument, minimum execution sizes, and matched volume, but not which fills were partial. They also learn which opposite orders satisfied each order's minimum size and bounds on its volume.\\

\noindent\textbf{All-for-One \cite{da2022all}\\}  
\noindent\emph{Design goal.} A ``no-operator'' dark pool in which the MPC is run by the venue's own traders or key stakeholders instead of a few trusted operators. The paper also shows that this scales to 100 parties at under a dollar per auction.

\noindent\emph{Entities.} Traders; up to 100 MPC parties drawn from the venue's traders or stakeholders (PP).

\noindent\emph{Workflow.} During each auction period, traders secret-share the volumes of their buy or sell orders, without limit prices, into separate buy and sell lists. At auction close, the parties compute total buy and sell volume. They then allocate the smaller total across both lists in order-number priority, allowing partial fills, and open every order's executed quantity. Trades execute at an external reference price. No settlement is specified.

\noindent\emph{Privacy profile.} Volumes and trader account IDs enter secret-shared, with security-with-abort under either a dishonest-majority (SPDZ) or an honest-majority (Shamir) threshold. The side and the number of orders per side are known through list membership. The output is each order's executed quantity, opened in plaintext to all parties, so everyone learns which orders filled fully, partially, or not at all.

\subsubsection{\S \; Encrypted Matching and Controlled-Disclosure Designs} \mbox{}\par

\medskip
\noindent\textbf{DI-PQ \cite{mazloom2023efficient}\\} 
\noindent\emph{Design goal.} A data-independent priority queue with polylogarithmic insertion and constant-time front access under MPC, used to build a continuous-double-auction dark pool whose per-order cost no longer grows linearly with the book.

\noindent\emph{Entities.} Traders; three MPC servers emulating the operator (PP).

\noindent\emph{Workflow.} A trader secret-shares an order's identifier, price, and volume with the servers, sending the asset symbol and side in the clear. Buy and sell orders sit in secret-shared max- and min-priority queues keyed by price. An incoming order is compared against the front of the opposite queue and matched while prices cross, for $\min(v_{\mathrm{buy}},v_{\mathrm{sell}})$ at the standing order's price. Its remainder is inserted, with a fully filled order inserted as a dummy at worst priority. The resulting matches are passed to an unspecified execution routine.

\noindent\emph{Privacy profile.} Sender, price, and volume stay secret-shared under an honest-majority, maliciously secure protocol; symbol and side are visible to the servers. Each price comparison is opened, and the matches are opened in plaintext to the servers: both identifiers, the volume, and the price. The servers therefore learn how many standing orders an order consumed, and that each consumed order except the last was fully filled. Whether the incoming order and the last standing order filled completely stays hidden, as do all unmatched prices and volumes.

\medskip
\noindent\textbf{Indifferential Privacy \cite{polychroniadou2025indifferential}\\}
\noindent\emph{Design goal.} A single-operator dark pool that hides order quantities from an untrusted auctioneer at near non-private speed, while still computing an exact maximum matching.

\noindent\emph{Entities.} Traders, a single dark-pool server (PP).

\noindent\emph{Workflow.} For an order of $x$ units, a trader submits $x+N$ unit nodes, where $N$ is drawn from a truncated geometric distribution. Each node carries its side and price in the clear and a commitment stating whether it is real or fake, with real nodes listed first. The server builds a bipartite graph linking buy and sell nodes whose prices cross. It repeatedly selects a polar-opposite pair: the most competitive price on one side against the least competitive price on the other. The owners of the pair open their commitments; if both nodes are real they are matched and removed. A fake node is discarded, and it reveals that the owner has no real nodes left. This continues until no compatible nodes remain.

\noindent\emph{Privacy profile.} Side, price, and trader identity are visible to the semi-honest server; an optional shuffler variant encrypts identities. Quantity is protected only by the noisy node count, under an $(\epsilon,\delta)$ indifferential-privacy guarantee. Each attempted pair is announced before matching, and each executed match reveals both identities and prices in plaintext. Once an order fills completely, its fake nodes are opened, so its full quantity becomes known.\\

\noindent\textbf{CorrDP \cite{chiang2023correlated}\\} 
\noindent\emph{Design goal.} Dark-pool mechanisms that are differentially private over whole trading rounds. They protect traders' inputs and also the correlations between their outputs, which in ordinary dark pools let a counterparty infer an honest trader's activity and front-run it.

\noindent\emph{Entities.} Traders, a liquidity provider, MPC servers emulating the trusted curator (PP).

\noindent\emph{Workflow.} Every trader participates in each round, secret-sharing a unit buy, sell, or dummy order under an identifier.
\begin{itemize}
  \item \emph{rDP-Volume-match.} The servers obliviously compute the maximum number of buy--sell pairs at an external reference price and assign matches. They then apply randomized response to each trade/no-trade bit and return it privately to its trader. Because randomized outcomes no longer balance, the liquidity provider covers the mismatch, and a noisy amount of its funds stays frozen until the privacy epoch ends.
  \item \emph{rDP-Double-auction.} Traders also submit their willingness at each discrete price. A clearing price is first chosen with the exponential mechanism, and then volume matching runs on the compatible orders.
\end{itemize}
No settlement is specified.

\noindent\emph{Privacy profile.} Orders, including their direction, stay secret-shared under an honest-majority assumption, and dummies hide whether a trader is active. Each trader receives only its own noisy trade/no-trade bit, and the double auction additionally publishes its clearing price. The liquidity provider sees only a noisy balance change, so the curator's outputs remain differentially private until the frozen funds are released.\\

\noindent\textbf{PrivacyDP \cite{asharov2020privacy}\\} 
\noindent\emph{Design goal.} A single-operator dark pool in which the operator matches orders without ever seeing their side, price, or volume, which stay hidden until a match is found.

\noindent\emph{Entities.} Traders, a single dark-pool operator (PP), a public exchange where matched orders are executed (Settle-mechanism).

\noindent\emph{Workflow.} Each trader encrypts its order (side, price, volume) under an FHE variant and sends it to the operator. The operator homomorphically checks the new order against the encrypted order book, obtaining an encrypted match result. A decryption sub-protocol, not detailed in the extended abstract, reveals whether a match exists. The matched orders are then decrypted, and the operator executes them on the public market, with special handling of partial fills so that the side stays hidden.

\noindent\emph{Privacy profile.} Side, price, and volume enter encrypted and stay hidden from the operator, though it knows who submitted each order. It learns in plaintext whether each incoming order matched. Matched orders are decrypted and executed on the public market, so their terms become public; unmatched orders stay encrypted.\\

\noindent\textbf{PET-Exchange \cite{hasselquist2023pet}\\}
\noindent\emph{Design goal.} A securities exchange running a continuous double auction with partial fills over an encrypted public order book, so that neither brokers nor the exchange see order prices or volumes.

\noindent\emph{Entities.} Clients (Trader), brokers (T-interm), exchange and matching engine (PP), an evaluator holding the decryption keys (PP).

\noindent\emph{Workflow.} Clients encrypt price and volume under the evaluator's per-instrument CKKS key and submit through a broker. The exchange inserts each order into the encrypted bid or ask book of its instrument by binary search. The evaluator resolves every comparison, in one of two ways:
\begin{itemize}
  \item \emph{Local.} The exchange computes an encrypted approximate comparison, and the evaluator decrypts only its $[0,1]$ result.
  \item \emph{Remote.} The exchange pads both values additively, and the evaluator decrypts and compares them.
\end{itemize}
In both cases the real query is hidden among random challenges whose known answers let the exchange detect a misbehaving evaluator. Matched orders fill the smaller volume, and the remainder is updated by homomorphic subtraction.

\noindent\emph{Privacy profile.} Price and volume stay encrypted, assuming an honest-but-curious exchange and an evaluator that does not collude with it. The instrument and side are visible from the book an order joins. The exchange learns every comparison result, and hence the price ranking of the book. The evaluator sees only context-free comparison values or padded differences. Executed orders become public, while unmatched prices and volumes stay encrypted.

\subsection{Blockchain-Native Asset Exchange Protocols}
\label{app:blockchain-protocols}

The blockchain-native asset-exchange protocols are grouped into \emph{Distributed Cryptographic Exchange Designs} and \emph{Co-ordinator-Assisted and Hybrid Private Exchange Systems}.

\subsubsection{\S \; Distributed Cryptographic Exchange Designs} \mbox{}\par

\medskip
\noindent\textbf{Renegade \cite{renegade}\\}
\noindent\emph{Design goal.} An on-chain dark pool in which no third party, including the block proposer, learns any balance, order, match, or trade history, removing extractable value and executing at midpoint prices.

\noindent\emph{Entities.} Traders, relayers holding the trader's match key (T-interm, PP), smart contract with a commitment tree and nullifier set (Settle-mechanism).

\noindent\emph{Workflow.} Each trader keeps a private wallet (balances, orders, fee approvals, keys) and places only a commitment to it on-chain. Every wallet update is proven valid in zero knowledge and nullifies the previous commitment. A trader delegates matching to a relayer, which gossips over a peer-to-peer network. Relayers handshake with commitments to an order, balance, and fee, plus a proof that they come from an unspent wallet. The two relayers then run a two-party maliciously secure MPC on the orders against public oracle midpoints. They generate a collaborative SNARK over the execution before opening the match and its proof together. The contract nullifies both wallets for matching and stores the matched amounts as encrypted notes, so each wallet can only settle next. Each trader then redeems its note into a new wallet with a validity proof.

\noindent\emph{Privacy profile.} Orders and balances never appear on-chain, but the relayer holding a trader's match key sees the whole wallet in plaintext. The public Gateway relayer cluster does the same for traders who use it. A handshake reveals no order attributes unless the trader attaches optional indications of interest, which selectively disclose them. The match tuple (amounts, side, midpoint price) is opened in plaintext only to the two counterparties. Settlement notes are encrypted under a key-private scheme, so observers learn only that some pair of traders matched, plus aggregate token flows.\\

\noindent\textbf{P2DEX \cite{baum2021p2dex}\\} 
\noindent\emph{\underline{Design goal:}}
A decentralized cross-chain cryptocurrency exchange immune to token theft and front-running. Servers match secret-shared orders under publicly verifiable MPC, and they hold deposited assets only under an all-server threshold signature.

\noindent\emph{Entities.} Clients (Trader), exchange servers that match and threshold-sign (PP), a smart contract with the underlying asset ledgers (Settle-mechanism), public verifiers of the MPC transcript (O-observ).

\noindent\emph{Workflow.} The servers post collateral to a smart contract and set up threshold-signature burner addresses on each ledger. A client transfers its tokens to its burner address, sends the servers a target address on the other ledger, and secret-shares a unit-quantity order with a hidden limit price. If any order or deposit is missing, the servers issue refund transactions instead. Otherwise they run a publicly verifiable MPC that obliviously sorts buy and sell limits. It matches the highest remaining buy against the lowest remaining sell while they cross, clearing each pair at the average of its two limits, and discards the rest. The servers then jointly sign payout transactions to the matched clients' target addresses.

\medskip
\noindent\colorbox{gray!12}
{\parbox{\dimexpr\linewidth-2\fboxsep\relax}{%
\textbf{Penalty mechanism.}
    A server that sends an invalid message, or withholds one after a complaint to the contract, is identified from the publicly verifiable transcript. A double-spend proof submitted by a client identifies all servers. Identified servers lose their deposits, which reimburse the affected clients and the honest servers.
}}\\

\noindent\emph{Privacy profile.} Limit prices enter secret-shared and stay hidden unless every server is corrupted. The on-chain deposit, however, publicly exposes the client's source address, the traded asset, and the amount. Unmatched orders are never revealed, and each pair's clearing price is opened only after all orders are processed. Payouts are ordinary on-chain transactions, so matched target addresses and exchanged amounts are public.\\

\noindent\textbf{Republic \cite{zhang2019republic}\\}
\noindent\emph{Design goal.} A decentralized dark pool for cryptocurrencies without a trusted operator. Nodes match secret-shared order fragments, and matched trades settle by cross-chain atomic swap between Bitcoin- and Ethereum-based assets.

\noindent\emph{Entities.} Traders, bonded order-matching nodes (PP), Registrar and Judge contracts on Ethereum, the underlying chains (Settle-mechanism).

\noindent\emph{Workflow.} Traders and nodes register with the Registrar by posting a REN bond; the Registrar partitions nodes into disjoint pools. A trader splits an order (pair, side, price, volume) into $n$ Shamir fragments, sends them to the designated pools, and pays an order fee that is refunded if the order expires. Nodes race to compute on fragments of order pairs, or larger tuples. The result is randomly scaled, so that recombining it reveals only whether the orders match. A match is registered on Ethereum, closing both orders, and the participating nodes share the fees. The counterparties then execute an atomic swap. The protocol cannot compel this swap; the bonds only incentivize it.

\medskip
\noindent\colorbox{gray!12}
{\parbox{\dimexpr\linewidth-2\fboxsep\relax}{%
\textbf{Penalty mechanism.}
A trader can challenge the matching nodes' computation before the Judge contract, with both sides staking their bonds. A node that cannot prove a truthful computation loses its bond, and a challenger whose claim is false loses its own.
}}\\

\noindent\emph{Privacy profile.} All order attributes enter secret-shared; an order stays hidden unless an adversary collects $n/2$ of its fragments. Controlling half the network therefore exposes about half the orders, without choosing which. Registration is public, and a trader's bond reveals an upper bound on its open orders. The match outcome is registered publicly on Ethereum, and the two counterparties learn some details of each other's orders. The atomic swap exposes assets, amounts, and addresses on the underlying chains.\\

\noindent\textbf{Rialto \cite{govindarajan2022privacy}\\}
\noindent\emph{Design goal.} A fully decentralized exchange marketplace, with no trusted custodian or operator, that keeps order rates and balances confidential and orders unlinkable to traders. It still provides front-running resilience, fair matching, and enough disclosure for price discovery.

\noindent\emph{Entities.} Traders, broker nodes running the MPC (PP), a blockchain with a smart contract holding committed balances, matching, and settling (Settle-mech).

\noindent\emph{Workflow.} Balances are stored on-chain as Pedersen commitments, and each order is for one unit at a limit rate. A trader additively shares its rate among the brokers and posts commitments to the shares. A buyer also proves in zero knowledge that its rate does not exceed its balance, and the contract escrows the funds homomorphically. When a timer expires or $N$ orders arrive, the brokers validate their shares against the commitments. They then run SortingMPC and publish a single list ordered by rate. The contract matches in the clear over this list, pairing each buyer with the first cheaper seller and carrying unmatched orders to later rounds. Each trader settles at its own rate: SettlementMPC outputs only the aggregate fee, and the contract updates balances homomorphically. Finally, the brokers re-randomize and obliviously shuffle all balance commitments through a Waksman network.

\noindent\emph{Privacy profile.} Rates and balances enter committed and secret-shared. Participation in a round and the side of each order are public, because buyers escrow. Rates stay hidden unless a threshold of brokers colludes, but the sorted list leaks the total ordering of all rates. Matched pairs are public by index, and each round opens the top-$K$ settlement rates and the aggregate fee in plaintext. The end-of-round shuffle unlinks accounts from their orders across rounds.

\subsubsection{\S \; Co-ordinator-Assisted and Hybrid Private Exchange Systems}  \mbox{}\par

\medskip
\noindent\textbf{SPCEX \cite{xie2024spcex}\\}
\noindent\emph{Design goal.} A cross-chain cryptocurrency exchange that guarantees atomicity and trader unlinkability without trusted hardware. It must also resist front-running by the exchanger, collusion, and Sybil attacks.

\noindent\emph{Entities.} Traders, an untrusted exchanger $E$ that matches orders for a fee (PP), a trusted supervisor $S$ that holds $E$'s collateral and assists comparisons without colluding with $E$ (PP), and the two ledgers with their escrow addresses (Settle-mechanism).

\noindent\emph{Workflow.} $E$ deposits collateral with $S$. Each trader mixes its tokens into an ephemeral address, pays $E$ a service fee and a transaction deposit, and escrows its tokens under a timelock. It then submits its limit, encrypted under a homomorphic re-encryption scheme, together with its side in the clear. To compare two limits, $E$ forms their encrypted difference and $S$ blinds it with a random value and sign, so $E$ learns only the sign of the comparison. $E$ sorts both sides and matches the highest buy against the lowest sell while they cross. For each matched pair, $E$ and $S$ jointly decrypt the sum of the two limits, and $E$ publishes the pair with its clearing price, the average. The counterparties settle through a hash-locked atomic swap mediated by $E$; unmatched or stalled traders reclaim their escrow after the timeout.\\

\noindent\colorbox{gray!12}
{\parbox{\dimexpr\linewidth-2\fboxsep\relax}{%
\textbf{Penalty mechanism.}
A trader that fails to complete a matched swap loses its transaction deposit to its counterparty. If $E$ colludes with a trader to double-spend or falsify a clearing price, the victim reports it to $S$ with evidence, and $S$ reimburses the victim from $E$'s collateral.
}}\\

\noindent\emph{Privacy profile.} Limits enter encrypted, but the side is in the clear and the escrow deposits are on-chain. $E$ learns only comparison signs, and hence the relative order of all limits. $S$ sees only blinded values. Matched pairs and their clearing prices are published in plaintext, while unmatched limits are never revealed. The swaps expose amounts and addresses on both ledgers, but mixing before submission and after receipt unlinks those addresses from traders.\\

\noindent\textbf{Deep Ocean \cite{francca2019deep}\\}
\noindent\emph{Design goal.} A blockchain-agnostic layer-two dark pool in which the operator keeps the order book without learning order sizes, and a trader reveals its size only once it knows the order will be completely filled.

\noindent\emph{Entities.} Traders, an honest-but-curious operator that registers users, relays messages, supplies the market price, and punishes deviations (T-interm), and an external settlement mechanism such as atomic swaps (Settle-mechanism).

\noindent\emph{Workflow.} Traders register with the operator, KYC-style, and send orders stating the assets to buy and sell and an optional limit price, withholding the size. The operator pairs compatible orders at the current market price, which traders check against public sources, and sends each side the other's public key; both confirm or abort without penalty. Relayed through the operator, the two traders compare their sizes. They encrypt them bitwise under a joint ElGamal key with zero-knowledge proofs, shuffle and blind the derived vectors, and jointly decrypt only which size is larger. The holder of the smaller order opens its size, the traders settle externally, and the remainder of the larger order returns to the book.

\medskip
\noindent\colorbox{gray!12}
{\parbox{\dimexpr\linewidth-2\fboxsep\relax}{%
\textbf{Penalty mechanism.}
Once both traders confirm a match, they must complete it. Because every message is signed and proven in zero knowledge, deviation is publicly provable, and the operator punishes the offender with a fine or expulsion from the pool.
}}\\

\noindent\emph{Privacy profile.} Identity, asset pair, and limit price are known to the operator in plaintext; only the size is hidden, and a sketched variant also hides limit prices. The comparison runs between the two traders and reveals only which size is larger. The smaller size is then opened to the counterparty and the operator. The larger order's size, and hence its unfilled remainder, stays hidden from everyone.\\ 

\noindent\textbf{COMMON \cite{garreta2023common}}\par
\noindent\emph{Design goal.} A decentralized order-book exchange built into a shielded token pool, so users trade without leaving the pool and keep their identities and order amounts hidden. Batch matching followed by a Dutch auction draws external liquidity and returns most MEV to users.

\noindent\emph{Entities.} Users (Trader), ORDER-BOOK and SWAP-ENGINE contracts with the shielded pool (Settle-mechanism), an off-chain Decryption Oracle run by a threshold committee, a TEE, or a trusted party (T-interm), market makers (PP), a price oracle.

\noindent\emph{Workflow.} A user spends a shielded note to post an order whose token pair and maximum price are public. The amount is an additively homomorphic ciphertext, the private part is committed as an order note, and a zk-SNARK ties them together. In each round's collect phase, orders whose maximum price meets an oracle-derived bound join the batch, and their encrypted amounts are summed per directed pair. The Decryption Oracle decrypts only these aggregates. The SWAP-ENGINE then matches opposite aggregates internally and sells the remainder to market makers in a Dutch auction, never below the batch bound. It logs one traded fraction and clearing price per batch. Users later claim their proceeds, or cancel, by proving in zero knowledge that the claim matches their order note, minting new notes into the pool.

\noindent\emph{Privacy profile.} Amounts and the link to the spending user enter hidden. Token pair, and hence direction, and the maximum price are public; dummy orders encrypting zero can mask direction. Only per-pair aggregates are decrypted, so individual amounts stay hidden unless all but one user in a batch is corrupted. The trade phase is public (aggregate volumes, clearing price, auction bids). Claims reveal only the pair and round.\\


\noindent\textbf{Singularity \cite{singularity,singularity_current}\\}
\noindent\emph{Design goal:}  To provide an on-chain institutional dark pool that hides order details from public observers while enabling private matching and verifiable on-chain settlement. 

\noindent\emph{Enitities:} Traders, BOOK nodes/ market makers (PP), relayers (T-interm), and smart contracts (Settle-mechanism). 

\noindent\emph{Workflow:} A trader first deposits assets into Singularity's shared multi-asset UTXO contract; the deposit becomes a private note whose commitment is appended to a Merkle tree with the root stored on-chain. For trading, the user generates an encrypted order intent locally and signs it off-chain, and a relayer broadcasts it as a meta-transaction over private RPC rather than the public mempool, so neither the trader's address nor its gas payment appears on chain. For matching, \cite{singularity} describes FHE over encrypted orders and MPC for key generation and revealing match outcomes, hence our \emph{Hybrid} (FHE+MPC) classification in Table~\ref{tab:matching-comparison}, though no source specifies the match computation in detail. Unmatched orders may alternatively be routed anonymously to the AMM for backup liquidity. Matched pairs are settled by the non-custodial contract, which verifies the proofs and executes the swap atomically, publishing only nullifiers and new note commitments.

\noindent\emph{Privacy profile:} Order parameters are encrypted before
submission to the BOOK, protecting their contents from public observers
during submission and matching. The BOOK operates on encrypted orders,
while the matching result is revealed to the relevant traders. The
available sources do not fully specify what information is exposed
during order discovery or subsequent trader-to-trader negotiation. Due to private notes, no order attributes are revealed during settlement; however public observers observe settlement activities such as state updates on blockchain \cite{singularity_current}.\\


\noindent\textbf{Tristero \cite{tristero,tristero_recent}\\}
\noindent\emph{Design goal:} To provide a decentralized private order-matching engine that separates confidential matching from settlement, keeping the order book encrypted inside trusted execution environments.

\noindent\emph{Entities:} Traders, a network of nodes (PP), sequencer (T-interm), and settlement contract (Settle-mech)
    
\noindent\emph{Workflow.} A trader salts its intent (address, asset, quantity), posts its hash on-chain as a commitment, and gossips the encrypted order to the nodes. The nodes validate it and agree on it through a sequencer-driven two-phase commit and PBFT consensus. Inside their enclaves, they keep an encrypted book and match incoming against resting orders. For each match, the nodes stage a transaction that the Ethereum contract checks against the committed intent hashes before transferring tokens. Nodes therefore never custody funds.

\noindent\emph{Privacy profile:} The asset, quantity, direction, and trader identity remain hidden until matching. After matching, however, the Ethereum settlement contract executes the transfers on-chain, exposing the executed trade and associated settlement activity to public observers, including the trader identity and quantity\cite{tristero,tristero_recent}.

\end{document}